\documentclass[fleqn,usenatbib]{mnras}

\usepackage{newtxtext,newtxmath}

\usepackage[T1]{fontenc}

\DeclareRobustCommand{\VAN}[3]{#2}
\let\VANthebibliography\thebibliography
\def\thebibliography{\DeclareRobustCommand{\VAN}[3]{##3}\VANthebibliography}

\usepackage{graphicx}	
\usepackage{amsmath}	
\usepackage{CJKutf8}

\title[The Schwarzschild Plunging Region]{The Plunging Region of a Thin Accretion Disc around a Schwarzschild Black Hole}

\author [Jake Rule et al.]{Jake Rule$^{1, 2, 3}$\thanks{E-mail:
jake.rule@physics.ox.ac.uk}, Andrew Mummery$^4$, Steven Balbus$^{1,4}$, James Stone$^2$, Lizhong Zhang (张力中) $^{2,5}$ \\
$^1$Oxford Astrophysics, Denys Wilkinson Building,  Keble Road, Oxford, OX1 3RH, United Kingdom \\
$^2$School of Natural Sciences, Institute for Advanced Study, 1 Einstein Drive, Princeton, NJ 08540, USA \\
$^3$Department of Astrophysical Sciences, Princeton University, Princeton, NJ 08544, USA \\
$^4$Oxford Theoretical Physics, Beecroft Building,  Clarendon Laboratory, Parks Road, Oxford, OX1 3PU, United Kingdom \\
$^5$Center for Computational Astrophysics, Flatiron Institute, New York, NY 10010, USA
}

\date{Accepted XXX. Received YYY; in original form ZZZ}

\pubyear{\the\year{}}

\begin{document}
\begin{CJK}{UTF8}{gbsn}
\label{firstpage}
\pagerange{\pageref{firstpage}--\pageref{lastpage}}
\maketitle
\end{CJK}
\begin{abstract}
A set of analytic solutions for the plunging region thermodynamics have been developed recently under the assumption that the fluid undergoes a gravity-dominated geodesic plunge into the black hole. We test this model against a dedicated 3D global GRMHD simulation of a thin accretion disc around a Schwarzschild black hole using the code {\tt AthenaK}. Provided that we account for non-adiabatic heating in the energetics, plausibly from grid-scale magnetic dissipation, we find an excellent agreement between the analytic model and the simulated quantities. These results are particularly important for existing and future electromagnetic black hole spin measurements, many of which do not to include the plunging fluid in their emission modelling. This exclusion typically stems from the assumption of a zero-stress boundary condition at the ISCO, forcing all thermodynamic quantities to vanish. Instead, we find a non-zero $\delta_\mathcal{J}\approx 5.3 \%$ drop in the angular momentum over the plunging region, which is consistent with both prior simulations and observations. We demonstrate that this stress is small enough for the dynamics of the fluid in the plunging region to be well-described by geodesic trajectories, yet large enough to cause measurable dissipation near to the ISCO - keeping thermodynamic quantities from vanishing. In the plunging region, constant $\alpha$-disc models are a physically inappropriate framework.
\end{abstract}




\section{Introduction}
From the supermassive black holes at the center of Active Galactic Nuclei to the stellar mass black holes found in X-ray binary systems, disc accretion of matter onto black holes is a prominent source of power in the Universe. Moreover, the emission of radiation close to a black hole `shines a light' on this strong gravity regime and is therefore an important tool for understanding General Relativity (GR), complementing the recent advent of gravitational wave detection.
\par
A feature of GR is the existence of a \emph{plunging region} between the Innermost Stable Circular Orbit (ISCO) and the event horizon of a black hole. In this region, no stable circular orbits exist, yet it is still a source of light and other material. The position of the ISCO and the size of the plunging region is strongly dependent on the black hole spin, denoted by the parameter $a = cJ/GM$, where $J$ is the angular momentum of the black hole and $M$ is its mass. Maximally prograde black holes ($a/M=1$), have no plunging region at all ($r_\text{ISCO} = r_\text{horizon} = GM/c^2$), whilst maximally retrograde black holes ($a/M=-1$) have the largest plunging region ($r_\text{ISCO} = 9\,r_\text{horizon} = 9\,GM/c^2$), a dependence which has been used as the basis for measuring the black hole spin.
\par
Unlike the mass, the spin of a black hole has no Newtonian counterpart and is most readily measured by probing the strong gravity regime. A common technique for measuring the spin of a black hole is based upon measuring the thermal continuum emission of its accretion disc \citep[][]{McClintock14, reynoldsObservationalConstraintsBlack2021}. Modelling this emission allows one to constrain the location of the ISCO and hence the spin. Using classical disc models, the most energetic emission in the spectrum originates from just outside of the ISCO. Since the liberated binding energy will increase as the ISCO moves closer to the black hole, measuring the energy of this edge allows one to determine the location of the ISCO.
\par
Generally, this approach and all other accretion-based spin measurement techniques, such as reflection spectroscopy, are dependent on the disc model used and on the prescription used to model the emission. Both of these are a source of systematic error for existing and future spin measurements \citep[][]{reynoldsObservationalConstraintsBlack2021}.
\par
The classical analytic model for steady disc accretion is the \cite{shakuraBlackHolesBinary1973} model, which describes a non-relativistic, radiatively efficient, optically thick and physically thin disc. This was then extended into full GR by \cite{novikovAstrophysicsBlackHoles1973} and \cite{pageDiskAccretionBlackHole1974}. To close these solutions, an assumption must be made about the boundary condition at the inner edge of the models validity (the point at which circular orbits become unstable). For black holes, the traditional assumption is that there is a zero stress boundary condition at the ISCO. At the time, the role of magnetic fields in accretion discs was uncertain, it was not until the discovery of magnetorotational instability (MRI) (\cite{balbusPowerfulLocalShear1991}) and its crucial role in the turbulent transport of angular momentum in discs (\cite{balbusInstabilityTurbulenceEnhanced1998}) that the possibility of a finite stress at the ISCO from the magnetic fields was seriously considered (\cite{krolikMagnetizedAccretionMarginally1999}, \cite{gammieEfficiencyMagnetizedThin1999} and \cite{agolMagneticStressMarginally2000}). Since then, simulations with increasing sophistication have confirmed the presence of a finite ISCO stress, including multiple global General Relativistic Magnetohydrodynamics (GRMHD) simulations (e.g. \cite{HawleyKrolik01, KrolikHawley02, shafeeThreeDimensionalSimulationsMagnetized2008}, \cite{pennaSimulationsMagnetizedDiscs2010}, \cite{nobleDependenceInnerAccretion2010}, \cite{zhuEyeStormLight2012}, \cite{schnittmanDiskEmissionMagnetohydrodynamic2016}, and \cite{Dhang25}, amongst others).
\par
One consequence of a zero stress boundary condition is that it forces all thermodynamic quantities to vanish at the ISCO and remain zero throughout the plunging region. Such zero stress models are routinely used for the aforementioned thermal continuum fitting method for measuring black hole spins, despite compelling numerical evidence that there is a finite ISCO stress. As a consequence, they completely ignore the plunging fluid \citep[this is a widespread practice, see e.g.,][for a review, or e.g., \citealt{Zhao21} for a recent example]{McClintock14}. \cite{mummeryAccretionInnermostStable2023} (hereafter MB23) derived a full set of analytic thermodynamic solutions for the plunging region flow which can, among other things, be used to include the radiation emitted by fluid in the plunging region into the models used for continuum fitting. Crucially, MB23 find that the most energetic radiation can originate from within the plunging region, an effect that is a potential source of systematic error for existing spin measurements using classical disc models. Radiation from the plunging fluid provides a possible explanation for X-ray spectra which did not fit well using the classical zero stress models (e.g.,  \cite{Fabian20,mummeryContinuumEmissionPlunging2024}, \cite{mummeryPlungingRegionEmission2024}).
\par
The calculation discussed here assumes that the plunging flow is well-modelled by a gravity-dominated geodesic plunge for thin and weakly magnetised discs. In practice, even when modelling thin, weakly magnetised discs it is unclear a priori if thermal and magnetic pressure gradients are dynamically important to the plunging flow. Targeted global GRMHD simulations of the main disc and plunging region are required to test the theory.
\par
While being the simplest case to model analytically, thin discs are demanding to simulate (e.g. \cite{liskaHAMRNewGPUaccelerated2022} for a recent discussion). This is because one must adequately resolve the vertical extent of the disc, which is by definition much 
smaller than the radial extent. Increased spatial resolution also requires shorter time steps to satisfy the Courant–Friedrichs–Lewy condition. Moreover, the diffusion timescale of thin discs scales as $t_{\text{diff}} \sim \left(h/r\right)^{-2}$, where $h/r$ is ratio of the scale height to the radius, meaning that for thinner discs the simulation must run for longer to reach an equilibrium. The plunging region itself also represents an intrinsically challenging regime for simulations. The flow is highly relativistic, with a quickly dropping thermal energy that must be recovered accurately despite being several orders of magnitude smaller than the evolved total fluid energy.
\par
In this paper we present and analyse an {\tt AthenaK} simulation of a thin accretion disc, building on previous work done using the same code for the thick disc case by \cite{mummeryThreedimensionalStructureBlack2024}. They found good agreement even in that, more extreme, case. To keep the disc thin, we introduce a cooling function (i.e. \cite{shafeeThreeDimensionalSimulationsMagnetized2008}, \cite{nobleDirectCalculationRadiative2009} and \cite{pennaSimulationsMagnetizedDiscs2010}), mimicking rest frame isotropic radiative losses and counteracting the unavoidable grid-scale numerical dissipation. Due to hardware and code advances over the last decade, this simulation represents a numerical improvement over the 2010-era simulations of the same kind (i.e. \cite{pennaSimulationsMagnetizedDiscs2010} and \cite{nobleDependenceInnerAccretion2010}). For instance, we do not artificially restrict the domain in the azimuthal direction, our grid is locally isotropic (with a cell aspect ratio of exactly $1:1:1$ everywhere), yet we achieve a better resolution in the mid-plane of the plunging region. 
\par
We find a remarkable agreement between the simulated quantities and the MB23 analytic model, provided that one accounts for observed non-adiabatic heating due to magnetic reconnection in the mid-plane of the plunging flow. We also find a small but non-zero ISCO stress ($\delta_\mathcal{J} \approx 5.3\%$), consistent with both previous simulations (\cite{pennaSimulationsMagnetizedDiscs2010}, \cite{nobleDependenceInnerAccretion2010}) and observations (\cite{mummeryContinuumEmissionPlunging2024}, \cite{mummeryPlungingRegionEmission2024}).
\par
The layout of the paper is as follows. In Section \ref{sec:Model}, we briefly review the MB23 model. In Section \ref{sec:Simulation}, we give an overview of our simulation setup before discussing the simulation results and comparing them with the model in Section \ref{sec:Results}. We draw conclusions in Section \ref{sec:Conclusion}.
\section{Plunging Region Accretion Model}
\label{sec:Model}
\subsection{Dynamics}
\label{subsec:modeldynamics}
The MB23 model assumes that within the ISCO, gravity is the dominant acceleration term in the ideal GRMHD Euler equation. Comparing the typical scales of the accelerations due to gravity, pressure and magnetic fields ($a_G$, $a_P$ and $a_B$ respectively) within the ISCO (which has radius $r_I$), they find,
\begin{equation*}
    a_G \sim \frac{c^2}{r_I} \gg a_P \sim \frac{c_S^2}{r_I} , a_B \sim \frac{v_A^2}{r_I} .
\end{equation*}
As MB23 point out, within the ISCO, gravity is the dominant acceleration wherever the sound speed, $c_S$, and the Alfven velocity, $v_A$, are sub-relativistic. For the pressure, they show that this assumption is self-consistent with the eventual thermodynamic solutions (for typical sub-Eddington choices of the accretion rate $\dot M$). Since the theory does not dynamically evolve the magnetic field, one can check for the self-consistency of this assumption only with full GRMHD simulations. MB23 posit that the Alfven velocity will be sub-relativistic except perhaps for when the disc is in a magnetically arrested disc state \citep{narayanMagneticallyArrestedDisk2003}.
\par
With gravitational accelerations dominant, the Euler equation is nothing more than the geodesic equation and so the fluid parcels follow a test particle geodesic plunge. The specific energy and specific angular momentum of a fluid parcel are given by the covariant components of the 4-velocity, $-U_0$ and $U_\phi$ respectively. For geodesic motion in the Kerr metric, these are both conserved. One can find plunging solutions to the geodesic equation with the same energy and angular momentum as a particle on a circular orbit at the ISCO (\cite{cunninghamEffectsRedshiftsFocusing1975}, \cite{mummeryInspiralsInnermostStable2022}). These have a radial 4-velocity,
\begin{equation}
    \label{eq:puregeo}
    U^r = -c \sqrt{\frac{2r_g}{3r_I}} \left(\frac{r_I}{r}-1\right)^\frac{3}{2} .
\end{equation}
\par
Formally, in this pure geodesic solution there is no radial velocity at the ISCO radius itself since the solution must reduce to a circular orbit there. In reality, if this model is to describe an accretion scenario there should be a non-zero radial velocity at the ISCO so that fluid elements actually pass over it. Therefore, $U_\phi$ (or $U_0$) cannot be exactly the same as for a circular orbit at the ISCO. Some angular momentum is passed back from the plunging flow to the main disc, and this angular momentum is assumed to source a small but non-zero stress at the ISCO. Without such a stress, one can also show from the relativistic equations for the exterior disc (\cite{balbusGeneralRelativisticThin2017}, \cite{mummeryAccretionInnermostStable2023}) that the radial velocity diverges at the ISCO, which is clearly unphysical. At the same time, the stress is assumed small enough that it does not lead to large departures from a geodesic plunge. The validity of such an assumption is one of the key tests performed in this work.

This angular momentum transport must conspire to produce a non-zero ISCO velocity, which can be approximately incorporated by including an offset velocity, $u_I$,
\begin{equation}
    \label{eq:offsetgeo}
    U^r = -c \sqrt{\frac{2r_g}{3r_I}} \left(\frac{r_I}{r}-1\right)^\frac{3}{2} - u_I .
\end{equation}
We can make this approximation because we expect that $u_I$ will be on the scale of the sound speed (\cite{mummeryDynamicsAccretionFlows2024}), which will be very small in comparison to the geodesic term once the fluid has been rapidly accelerated by gravity to near the speed of light. 

\subsection{Thermodynamics}
\label{subsec:modelthermodynamics}
With a constant mass accretion rate $\dot{M} = 2\pi r \Sigma (-U^r)$, one can use the radial velocity (Eq.\,\ref{eq:offsetgeo}) to find the surface density as a function of radius $\Sigma(r)$,
\begin{equation}
    \label{eq:massconsv}
    \Sigma(r) = \frac{\dot{M}}{2 \pi r (-U^r)}.
\end{equation}
It is important to point out that in a zero-stress model, $u_I\to\infty$ and the surface density vanishes, along with all other thermodynamic variables. For a non-vanishing model of the plunging region thermodynamics, it is therefore crucial to have a non-zero ISCO stress and a finite $u_I$.
\par
MB23 assume that the plunging flow is well-approximated by being adiabatic. This approximation, which we will test in this work, is likely to be accurate if the energy loss/gain timescales are longer than the free-fall time of the flow, which we note is short (of order the light crossing time) within the ISCO: $t_{\rm ff}\sim r_I/c$. With an ideal gas equation of state,
\begin{equation*}
    u_g = \frac{P}{\gamma-1},
\end{equation*}
the conservation of entropy leads to,
\begin{equation}
    P=K \rho^\gamma.
\end{equation}
Where $\gamma$ is the adiabatic index and $K=P\rho^{-\gamma}$, which is closely related to the specific entropy $s\propto\log(K)$. In the above expression $u_g$ is the internal energy, $P$ is the gas pressure and $\rho$ is the rest mass density of the fluid. The rest mass density can be found from the surface density,
\begin{equation*}
    \rho = \frac{\Sigma}{H},
\end{equation*}
where $H$ is the scale height of the disc. MB23 then determine the scale height as a function of the pressure and density, assuming that a vertical hydrostatic equilibrium is maintained and that the fluid follows a geodesic plunge. This is another assumption which we shall test in this work. This is sufficient to close the set of equations. Ultimately they find that,
\begin{align}
    &\frac{\Sigma}{\Sigma_I} = \left(\frac{r_I}{r}\right)\left[\epsilon^{-1} \left(\frac{r_I}{r}-1\right)^\frac{3}{2}+1\right]^{-1}
    \label{eq:surfdensitymodel}
    \\
    &\frac{\rho}{\rho_I} = \left(\frac{r_I}{r}\right)^\frac{6}{\gamma+1}\left(\frac{K}{K_I}\right)^{\frac{-1}{1+\gamma}}\left[\epsilon^{-1} \left(\frac{r_I}{r}-1\right)^\frac{3}{2}+1\right]^{-\left(\frac{2}{\gamma+1}\right)}
    \label{eq:densitymodel}
    \\
    &\frac{P}{P_I} = \left(\frac{r_I}{r}\right)^\frac{6\gamma}{\gamma+1}\left(\frac{K}{K_I}\right)^{\frac{1}{1+\gamma}}\left[\epsilon^{-1} \left(\frac{r_I}{r}-1\right)^\frac{3}{2}+1\right]^{-\left(\frac{2\gamma}{\gamma+1}\right)}
    \label{eq:pressuremodel}
    \\
    &\frac{T_c}{T_{c,I}} = \left(\frac{r_I}{r}\right)^\frac{6(\gamma-1)}{\gamma+1}\left(\frac{K}{K_I}\right)^{\frac{2}{1+\gamma}}\left[\epsilon^{-1} \left(\frac{r_I}{r}-1\right)^\frac{3}{2}+1\right]^{-\left(\frac{2(\gamma-1)}{\gamma+1}\right)}
    \label{eq:tempmodel}
    \\
    &\frac{H}{H_I} = \left(\frac{r_I}{r}\right)^{-\left(\frac{5-\gamma}{\gamma+1}\right)}\left(\frac{K}{K_I}\right)^{\frac{1}{1+\gamma}}\left[\epsilon^{-1} \left(\frac{r_I}{r}-1\right)^\frac{3}{2}+1\right]^{-\left(\frac{\gamma-1}{\gamma+1}\right)}
    \label{eq:heightmodel}
\end{align}
Where $\epsilon$ is defined by
\begin{equation}
    \label{eq:epsilon}
    \epsilon \equiv \frac{u_I}{c} \sqrt{\frac{3r_I}{2r_g}},
\end{equation}
which should be viewed as a parameter of the model. We have explicitly kept the dependence on the quantity $K=P\rho^{-\gamma}$, so that any effects of turbulent heating/radiative cooling can be accounted for. We have also assumed that the system is gas pressure dominated, so that the central temperature, $T_c \propto P/\rho$. Note the self similar nature of these solutions for fixed $\epsilon$. The scaled solutions are exactly the same for all black hole spins and depend only on the radius in units of the ISCO.

\section{GRMHD Simulation}
\label{sec:Simulation}
To simulate a thin accretion disc around a black hole, we work within the ideal GRMHD framework, solving the fluid equations using the {\tt AthenaK} (\cite{stoneAthenaKPerformancePortableVersion2024}) code, which is based upon a previous code called {\tt ATHENA\,++} (\cite{stoneAthenaAdaptiveMesh2020}, \cite{whiteExtensionAthenaCode2016}). {\tt AthenaK} has been written using the Kokkos performance portability library (\cite{trottKokkosEcoSystemComprehensive2021}), which aims to be hardware agnostic and crucially means that {\tt AthenaK} can be run on GPU based systems as well as CPU based ones, leading to a considerable speed-up and the ability to run higher resolution studies.
\par
{\tt AthenaK} uses Cartesian Kerr-Schild (CKS) coordinates ($t,x,y,z$) (\cite{kerrNewClassVacuum1965}). It is, however, useful to analyse our data in Spherical Kerr-Schild (SKS) coordinates ($t,r,\theta,\phi$). These are defined from CKS coordinates in the following way:
\begin{align}
    x = (r \cos\phi - a \sin\phi)\sin\theta \, , \\
    y = (r \sin\phi + a \cos\phi)\sin\theta \, ,\\
    z = r \cos\theta \, ,
\end{align}
where $a$ is the black hole spin parameter and the time coordinate is unchanged. SKS coordinates transform to the well-known Boyer-Lindquist coordinates ($t_\text{BL},r_\text{BL},\theta_\text{BL},\phi_\text{BL}$) (\cite{boyerMaximalAnalyticExtension1967}) in the following way:
\begin{align}
    {\rm d}t_\text{BL} = {\rm d}t - \frac{2Mr\, {\rm d}r}{r^2-2Mr+a^2} \,,\\
    {\rm d}r_\text{BL} = {\rm d}r \,, \\
    {\rm d}\theta_\text{BL} = {\rm d}\theta \, , \\
    {\rm d}\phi_\text{BL} = {\rm d}\phi - \frac{a\, {\rm d}r}{r^2-2Mr+a^2} \,.\\
\end{align}
Since the radial coordinates are identical, radial profiles in SKS coordinates are identical to those in Boyer-Lindquist coordinates. In the following, all distances are written in units of $r_g = GM/c^2$ and all times in units of $t_g = GM/c^3$, scaling linearly with the black hole mass, $M$. Note that for a black hole, $-r_g\leq a \leq r_g$, and we take $a=0$ in this work. In code units we then set $GM=1=c$.
\par
A major challenge when simulating thin discs is having sufficient resolution to resolve the scale-height of the disc. In our simulation, this lies in the mid-plane ($z=0$) of the domain. We use a static mesh refinement (SMR) scheme to achieve this, which is detailed in Table \ref{tab:refinement}. A $1:1:1$ cell aspect ratio is maintained at all times. In the mid-plane, there are a total of 1280 grid cells between $x,y =0$ and $x,y=64$. Most relevantly, in the innermost regions the minimum cell spacing is $0.025 \, r_g/\text{cell}$. A lower resolution simulation is presented in Appendix \ref{appA}.
\begin{table}
\centering
\begin{tabular}{llll}
\hline
Refinement Level & $x,y \in $      & $z \in$ & Cell Spacing     \\ \hline
0     & $[-64,64]$ & $[-32,32]$ & $0.2 \, r_g/\text{cell}$ \\
1     & $[-64,64]$ & $[-16,16]$ &$0.1\,r_g/\text{cell}$ \\
2     & $[-32,32]$ & $[-8,8]$  & $0.05\,r_g/\text{cell}$ \\
3     & $[-16,16]$ & $[-4,4]$  &$0.025\,r_g/\text{cell}$ \\ \hline
\end{tabular}
\caption{
Static mesh refinement configuration. The spatial resolution doubles at each refinement level.
}
\label{tab:refinement}
\end{table}
\par
We use a second-order accurate Runge-Kutta (RK2) time evolution scheme, with a Courant–Friedrichs–Lewy number of $0.3$. For spatial reconstruction, we use a fourth-order accurate Piecewise-Parabolic Method (PPM4) (\cite{colellaPiecewiseParabolicMethod1984}). Finally, we use a Harten–Lax–van Leer+Einfeldt (HLLE) approximate Riemann solver (\cite{hartenUpstreamDifferencingGodunovType1983}, \cite{einfeldtGodunovTypeMethodsGas1988}). {\tt AthenaK} uses the total energy based conserved to primitive variable inversion scheme from \cite{kastaunRobustRecoveryPrimitive2021}.
\par
\begin{figure}
    \centering
    \includegraphics[width=\linewidth]{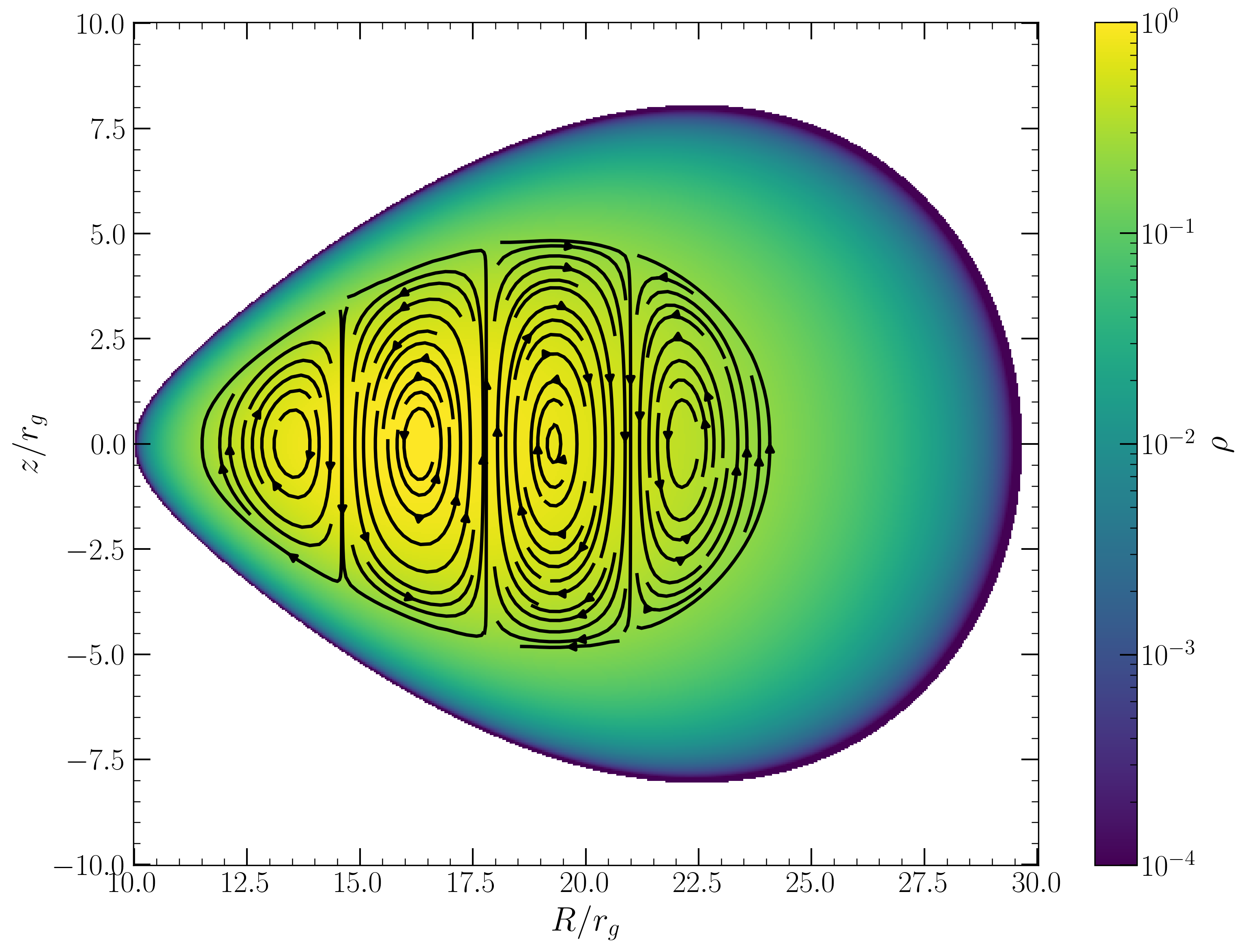}
    \caption{A cross-section of the initial Fishbone-Moncrief torus in the $R$-$z$ plane ($R=r\sin\theta$ is the cylindrical radius). The torus extends symmetrically in the $\phi$ direction. The density is shown in colour, whilst the black lines mark the initial four-loop poloidal magnetic field. The torus has an inner edge at $r_\text{in} = 10 \, r_g$ and pressure maximum at $r_\text{max} = 16 \, r_g$ ($z=0)$. These fix the outer edge to $r_\text{out} = 29.7 \, r_g$.}
    \label{fig:FM_Torus}
\end{figure}

To initialise the disc, we start with a \cite{fishboneRelativisticFluidDisks1976} torus, a hydrostatic equilibrium solution in the Kerr spacetime. The torus is parametrised by an inner edge and a maximum pressure radius (along the $z=0$ direction), which we set to $r_\text{in} = 10 \, r_g$ and $r_\text{max} = 16 \, r_g$. The density is normalised in code units so that it has a value of $\rho_\text{max}=1$ at the density/pressure maximum, as shown in Fig.\,\ref{fig:FM_Torus}. The density and pressure are related by $P=K_\text{torus}\rho^\gamma$, where $K_\text{torus}$ is a constant specified by the size of the torus. We set $\gamma = 13/9$, which is intended to reflect the relativistic electrons and non-relativistic ions in the accreting plasma (see e.g., \cite{whiteResolutionStudyMagnetically2019}).
\par
For numerical stability, we require the density and pressure to be above floors which we set globally at $\rho_f = 1\times10^{-8}$ and $p_f = 3.33 \times 10^{-11}$ in code units, so that they are several orders of magnitude smaller than the smallest densities and pressures that naturally occur in the disc. We also have an excise region within the black hole ($r\leq 1$) where the density and pressure are held at their floor values and the velocities are set to zero. For the Schwarzchild black hole considered in this paper, this is well within the event horizon, $r_+ = 2$.
\par
While in principle {\tt AthenaK} is solving the equations of {\it ideal} GRMHD, there is unavoidable numerical dissipation when energy turbulently cascades down to the grid scale. This dissipation heats the disc, making it thicker. In reality, the thin discs which we aim to simulate are 
efficiently cooled by radiation. Unfortunately, fully simulating the radiation physics is computationally expensive. Therefore, we use an ad-hoc prescription to model these radiative losses and keep the disc thin. Following the example of \cite{shafeeThreeDimensionalSimulationsMagnetized2008}, \cite{pennaSimulationsMagnetizedDiscs2010}, \cite{nobleDirectCalculationRadiative2009} and \cite{nobleDependenceInnerAccretion2010}, the conservation of stress-energy ($T_\mu^\alpha)$ can be modified to include a source term,
\begin{equation}
    \nabla_\alpha\left(T_{\mu}^{\alpha}\right) = S_\mu,
\end{equation}
where $\nabla_\alpha$ represents the covariant derivative and $S_\mu$ is given by,
\begin{equation}
    S_\mu = \mathcal{L} U_\mu.
\end{equation}
$\mathcal{L}$ is a cooling function which we borrow from \cite{pennaSimulationsMagnetizedDiscs2010},
\begin{equation}
    \mathcal{L} = - u_g \frac{\ln{\left(K/K_t\right)}}{\tau_\text{cool}},
\end{equation}
where $K=P\rho^{-\gamma}$ represents the specific entropy and $K_t$ is a target entropy constant below which there is no cooling. The cooling timescale is set equal to $\tau_\text{cool} = 2 \pi \left(a + R^{3/2}r_g^{-1/2}\right)$, which is the coordinate time taken for a single circular mid-plane orbit at the cylindrical radius $R=r\sin\theta$ ($a$ is the black hole spin parameter). This function aims to remove the entropy generated by the grid scale dissipation. It also has the benefit of not including any prior prescription for the scale height, ensuring that the disc settles naturally and without a prescribed $H(r)$ profile. This is important for our purposes, as we seek to test the MB23 scale height prescription from first principles.
\par
We initialise the torus with a weak magnetic field, from which the magneto-rotational instability (MRI, \cite{balbusPowerfulLocalShear1991}) develops. We normalise the magnetic field so that $\beta_\text{min}=\min(u_g/u_\text{mag})_\text{torus}=100$, which is the minimum ratio of the thermal energy ($u_g$) to the electromagnetic energy ($u_\text{mag}$) in the initial torus. The MRI destabilises the hydrostatic equilibrium after a few orbits, initiating a turbulent phase of accretion onto the central black hole. It has been found that the choice of initial magnetic field geometry can significantly alter the ultimate steady state (e.g. \cite{pennaSimulationsMagnetizedDiscs2010}). Since we wish to end up with a weakly magnetised disc, we use a four-loop purely poloidal field configuration as our magnetic field initial condition, as demonstrated in Fig.\,\ref{fig:FM_Torus}. To test if we are resolving the MRI, we measure fluid frame MRI quality factors in the vertical and azimuthal directions (as defined by \cite{nobleDependenceInnerAccretion2010} \& \cite{hawleyASSESSINGQUANTITATIVERESULTS2011}). We find that $Q_{\text{MRI},\,\theta}\gtrsim 10$ and $Q_{\text{MRI},\,\phi}\gtrsim 50$ in the main bulk of the disc at late times. These surpass the standard set by \cite{hawleyASSESSINGQUANTITATIVERESULTS2011}.
\par
Where there is no initial torus, we specify an initial background density that falls off with radius, $\rho = 10^{-2}\times r^{-3/2}$. As before, the pressure is related via an ideal gas equation of state, $P=K_t\rho^\gamma$, where $K_t$ is the target specific entropy of the cooling function. This ensures that the background is not initially cooled.
\par
During the course of our simulations, we have found that in the plunging region the numerical scheme which we have outlined above can, on occasion, become inaccurate within individual grid cells due to 
unavoidable round-off errors. Within the plunging region, the fluid is rapidly accelerated, reaching radial velocities (near to the horizon) of almost the speed of light. For thin discs with weak thermal support, the thermal component of the total energy can become considerably smaller than the kinetic component. Since the kinetic and thermal components are evolved together (as part of the stress-energy tensor), any error from the kinetic component (i.e. from round-off error) is therefore 
amplified, in relative terms, for the thermal component. The fluid-frame thermal energy that is recovered from the stress-energy tensor can therefore become inaccurate. 
To stabilise against this difficulty, one could impose a high floor so that the thermal energy does not become too small. However, since the purpose of this work is to study the thermodynamics of the plunging fluid, this is not a viable option. Instead, for cells that develop an 
inaccurate thermal energy, we replace the flux-calculated value by the average of well-behaved neighbouring cells. Whilst this is better than resorting to the floor, it still leads to errors on individual grid cells from snapshot to snapshot. 
\par
Nonetheless, we find that the \emph{averaged} radial profiles for thermal quantities (such as the pressure or gas temperature) are robust as a function of time (they do not exhibit large temporal variances, see \S\,\ref{sec:Results}). This indicates that whilst the scheme may resort to the local cell average in specific problematic cells, it still recovers the bulk radial behaviour of the flow. We believe that this therefore represents a good benchmark against which to test the MB23 model.
\section{Results}
\label{sec:Results}
In this paper, we present results from a simulation with a Schwarzschild ($a/M=0$) black hole. We ran the simulation for $20,000$\,$t_g$ (about 217 circular orbits at the ISCO). In this simulation, we set $K_t=0.001$ and the disc settled to a scale-height of $h/r \sim 0.06$. 
\par
\begin{figure}
    \centering
    \includegraphics[width=\linewidth]{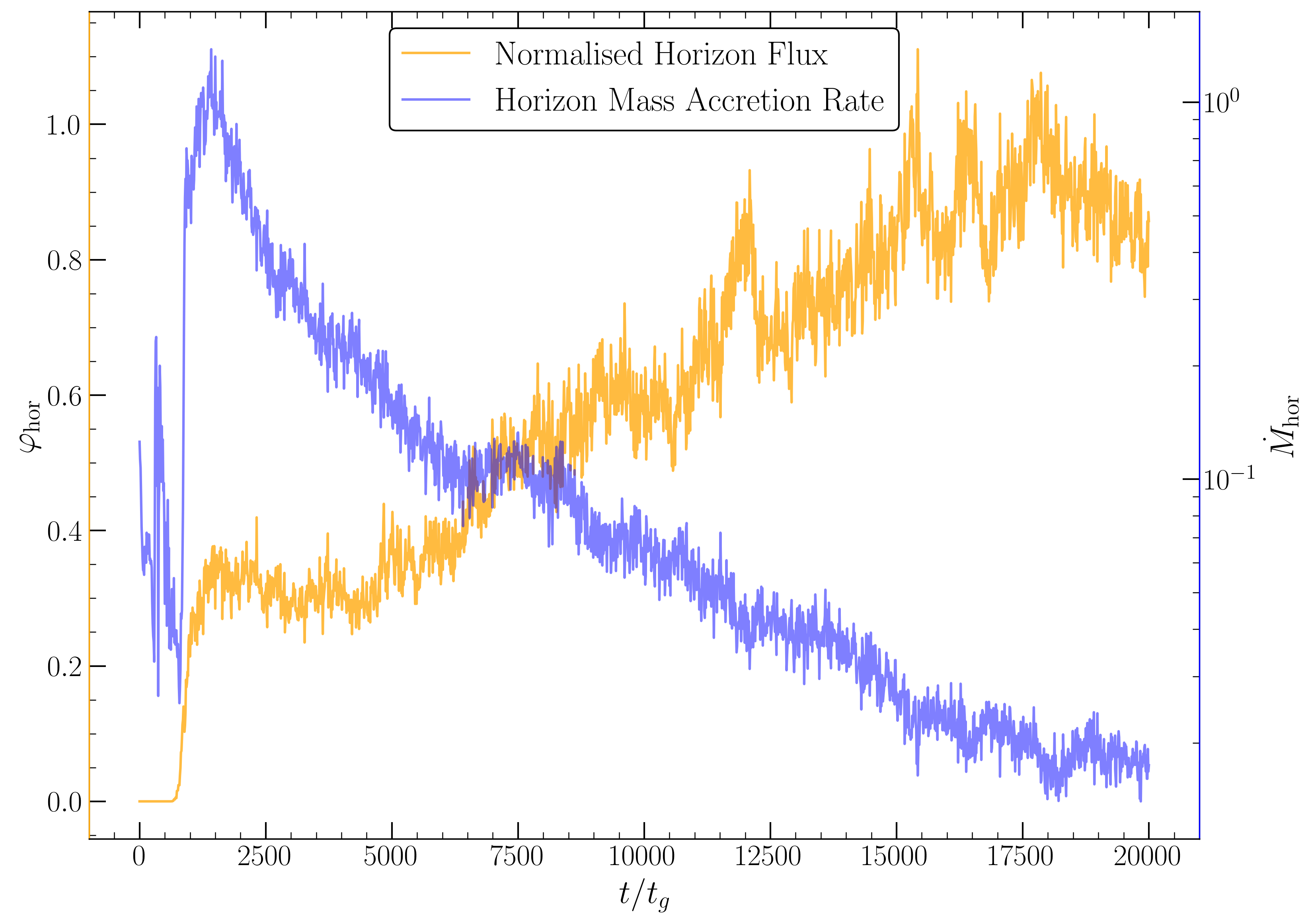}
    \caption{
    The time evolution of the mass accretion rate (right, logarithmic scale) at the event horizon $\dot{M}_\text{hor}$ (in blue), and a normalised magnetic flux at the horizon (left, linear scale), $\varphi_\text{hor} \equiv \Phi_\text{hor}\left(\dot{M}_\text{hor}r_g^2c\right)^{-1/2}$ (in orange). 
    }
    \label{fig:MdotandPhi}
\end{figure}

Fig. \ref{fig:MdotandPhi} shows the time evolution of the mass accretion rate at the horizon,
\begin{equation}
    \dot{M}_\text{hor} = -\int_0^{\pi}\int_0^{2\pi} \rho U^r \sqrt{g} d\phi d\theta,
\end{equation}
and a normalised magnetic horizon flux, $\varphi_\text{hor} \equiv \Phi_\text{hor}\left(\dot{M}_\text{hor}r_g^2c\right)^{-1/2}$ where,
\begin{equation}
    \Phi_\text{hor} = 
    \frac{1}{2} \int_0^{\pi}\int_0^{2\pi}  \lvert B^r \rvert \sqrt{g} d\phi d\theta,
\end{equation}
is the hemispherical magnetic horizon flux due to the factor of $1/2$ (following \cite{tchekhovskoyEfficientGenerationJets2011}). The mass accretion rate has a brief spike at early times, but begins to settle once the initial torus has been turbulently disrupted. By $t\sim15,000\,t_g$, the mass inflow is steady enough to produce time averaged profiles. The normalised magnetic flux is a standard test for magnetically arrested discs (MADs). Recent literature (e.g. \cite{narayanJetsMagneticallyArrested2022}, \cite{scepiMagneticSupportWinddriven2024}) suggests that MADs typically have a normalised flux of about $\varphi_\text{hor}\sim50$. Although the flux rises throughout the simulation, it remains firmly within the weakly magnetised regime.
\par
For the purposes of constructing radial profiles to test the  MB23 model, we require a steady-state averaging window. We use the window where $15,000< t/t_g < 20,000$, since there is a more-or-less steady mass inflow and a stable, weak magnetic horizon flux.
\par
We define our averages in the following way:
\begin{equation}
    \langle q_\text{mid} \rangle_{t \phi} = \frac{1}{t_f-t_i} \int_{t_i}^{t_f} \left[ \frac{\int_0^{2\pi}  q_\text{mid} \sqrt{g_{[t \phi]}} d\phi} {\int_0^{2\pi}  \sqrt{g_{[t \phi]}} d\phi}\right]dt,
\end{equation}
is the mid-plane temporal and azimuthal average of $q$, where $g_{[t \phi]}$ is the determinant of the $t \phi$ sub-metric, or equivalently the determinant of the induced metric  on hyper-surfaces of constant $r$ and $\theta$.
\begin{equation}
    \langle q \rangle_{t \phi \theta} = \frac{1}{t_f-t_i}\int_{t_i}^{t_f} \left[ \frac{\int_0^{2\pi}\int_{\theta_u}^{\theta_l}  q \sqrt{g_{[t \phi \theta]}} d\phi d\theta} {\int_0^{2\pi}\int_{\theta_u}^{\theta_l}  \sqrt{g_{[t \phi \theta]}} d\phi d\theta}\right]dt,
\end{equation}
is the temporal, azimuthal and vertical average of $q$, where $g_{[t \phi \theta]}$ is the determinant of the $t \phi \theta$ sub-metric. In the vertical direction we integrate from $\theta_u = 76^\circ$ to $\theta_l=104^\circ$, so as to not include non-disc material. This is also the most refined region centred about the mid-plane. Finally,
\begin{equation}
    \langle q \rangle_{\rho, \, t \phi \theta} = \frac{1}{t_f-t_i}\int_{t_i}^{t_f}\left[\frac{\int_0^{2\pi}\int_{\theta_u}^{\theta_l} q \rho \sqrt{g_{[t \phi \theta]}} d\phi d\theta} { \int_0^{2\pi}\int_{\theta_u}^{\theta_l} \rho \sqrt{g_{[t \phi \theta]}} d\phi d\theta}\right]dt,
\end{equation}
is the density weighted temporal, azimuthal and vertical average of $q$. Density weighting is useful for some physical quantities, where (for example) we are only interested in the quantity evaluated in the bulk of the disc flow, not including the background. We also define the azimuthally and temporally averaged surface density:
\begin{equation}
    \langle \Sigma \rangle_{t \phi} = \frac{1}{t_f-t_i}\int_{t_i}^{t_f}\left[\frac{\int_0^{2\pi}\int_{\theta_u}^{\theta_l}  \rho \sqrt{g_{[t \phi \theta]}} d\phi d\theta} { \int_0^{2\pi}\sqrt{g_{[t \phi]}} d\phi}\right]dt .
\end{equation}
To quantify the spread of the averaged radial profiles, we further define the following \emph{temporal} variances:
\begin{align}
    &\mathrm{Var}_{t \phi}( q_\text{mid} ) = \frac{1}{t_f-t_i} \int_{t_i}^{t_f} \left[ \frac{\int_0^{2\pi}  q_\text{mid} \sqrt{g_{[t \phi]}} d\phi} {\int_0^{2\pi}  \sqrt{g_{[t \phi]}} d\phi}\right]^2  \,{\rm d}t \, \nonumber\\ &\quad\quad\quad\quad\quad\quad\quad\quad\quad\quad\quad\quad\quad\quad\quad\quad\quad\quad-\,\langle q_\text{mid} \rangle_{t \phi}^2,
    \\
    &\mathrm{Var}_{t \phi \theta}(q) = \frac{1}{t_f-t_i}\int_{t_i}^{t_f} \left[ \frac{\int_0^{2\pi}\int_{\theta_u}^{\theta_l}  q \sqrt{g_{[t \phi \theta]}} d\phi d\theta} {\int_0^{2\pi}\int_{\theta_u}^{\theta_l}  \sqrt{g_{[t \phi \theta]}} d\phi d\theta}\right]^2 \,{\rm d}t \nonumber \\  &\quad\quad\quad\quad\quad\quad\quad\quad\quad\quad\quad\quad\quad\quad\quad\quad\quad\quad \,-\,\langle q \rangle_{t \phi \theta}^2,
    \\
    &\mathrm{Var}_{\rho, \, t \phi \theta}(q)= \frac{1}{t_f-t_i}\int_{t_i}^{t_f}\left[\frac{\int_0^{2\pi}\int_{\theta_u}^{\theta_l} q \rho \sqrt{g_{[t \phi \theta]}} d\phi d\theta} { \int_0^{2\pi}\int_{\theta_u}^{\theta_l} \rho \sqrt{g_{[t \phi \theta]}} d\phi d\theta}\right]^2\, {\rm d}t\nonumber \\ 
    &\quad\quad\quad\quad\quad\quad\quad\quad\quad\quad\quad\quad\quad\quad\quad\quad\quad\quad\,-\, \langle q \rangle_{\rho, \, t \phi \theta}^2.
    \\
    &\mathrm{Var}_{t \phi}(\Sigma)= \frac{1}{t_f-t_i}\int_{t_i}^{t_f}\left[\frac{\int_0^{2\pi}\int_{\theta_u}^{\theta_l} \rho \sqrt{g_{[t \phi \theta]}} d\phi d\theta} { \int_0^{2\pi} \sqrt{g_{[t \phi]}} d\phi}\right]^2\, {\rm d}t\nonumber \\ 
    &\quad\quad\quad\quad\quad\quad\quad\quad\quad\quad\quad\quad\quad\quad\quad\quad\quad\quad\,-\, \langle \Sigma \rangle_{t \phi}^2.
\end{align}
\par
\begin{figure*}
    \centering
    \includegraphics[width=0.49\linewidth]{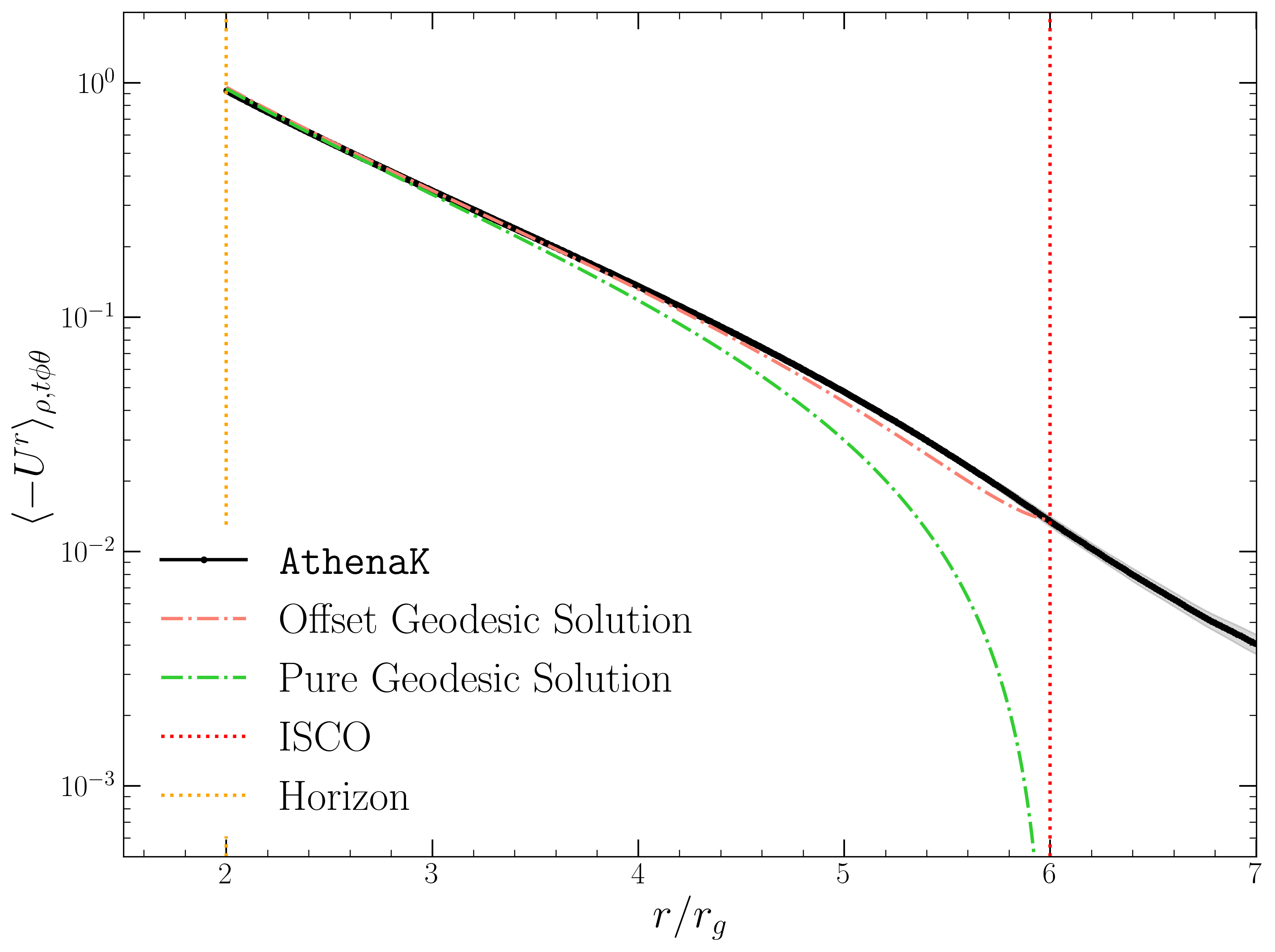}
    \includegraphics[width=0.49\linewidth]{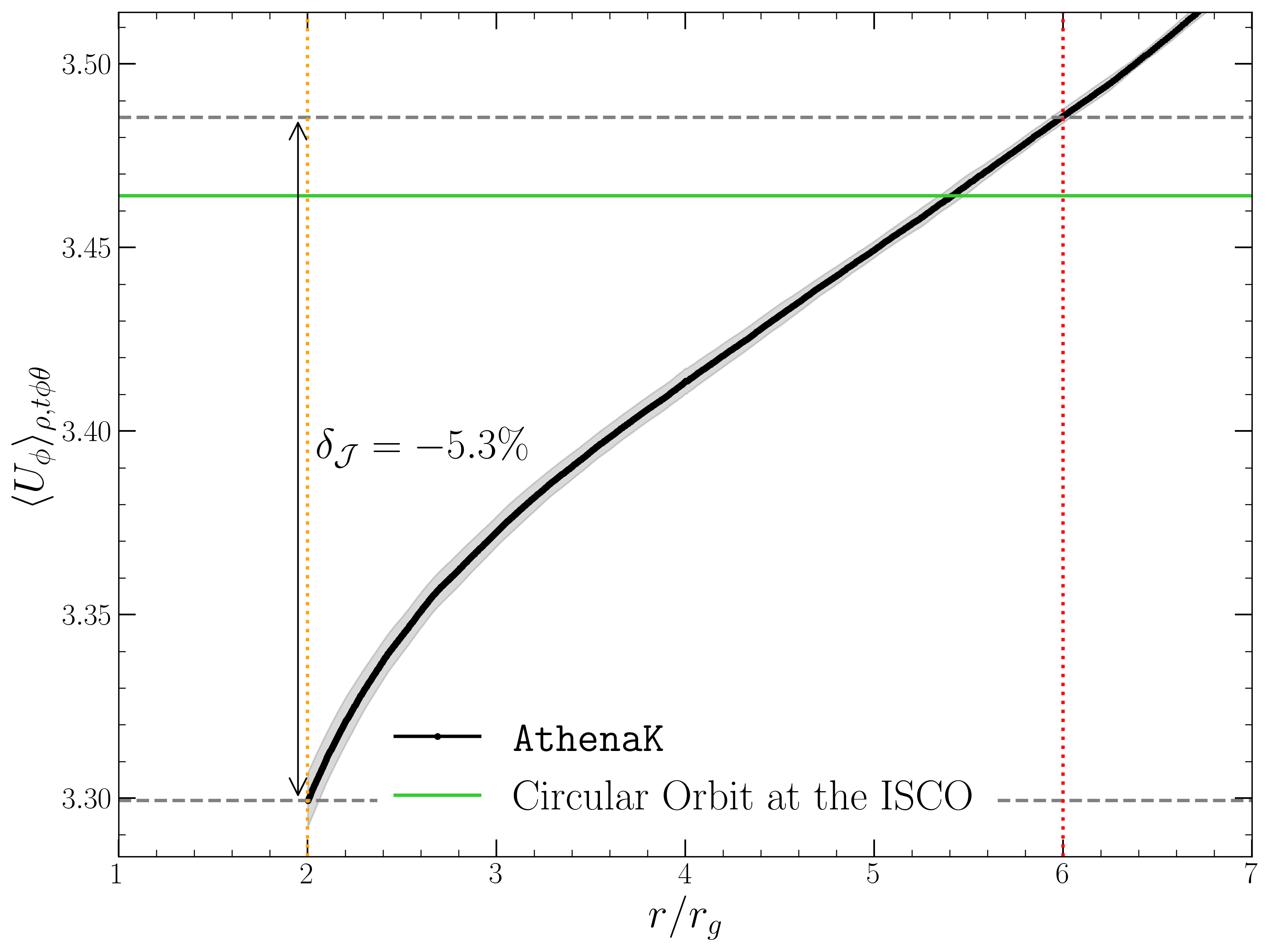}
    \caption{The plunging region radial 4-velocity ($U^r)$ (left) and angular momentum ($U_\phi)$ (right) profiles. The black dots in the upper left and right-hand panels show the density weighted, temporally ($[15,20]\text{k}t_g$), azimuthally ($[0,2\pi]$) and vertically ($[76^\circ,104^\circ]$) averaged simulated quantities, $\langle-U^r\rangle_{\rho, t \phi \theta}$ and $\langle U_\phi \rangle_{\rho, t \phi \theta}$. The shaded region represents $\pm1\sigma$ standard deviation $\sigma=\sqrt{\mathrm{Var}_{t\phi\theta}(-U^r)}$ and $\sigma=\sqrt{\mathrm{Var}_{t\phi\theta}(U_\phi)}$ respectively. The dot-dash lines in the left-hand panel are geodesic solutions. In orange is the offset geodesic from Eq.\,\ref{eq:offsetgeo} and in green is the pure geodesic from Eq.\,\ref{eq:puregeo}. The horizontal solid green line in the right-hand plot is the angular momentum corresponding to a circular orbit at the ISCO. There is a $\delta_\mathcal{J} = -5.3\%$ drop in the average angular momentum over the plunge.
    }
    \label{fig:urfit}
\end{figure*}

The left-hand panel of Fig.\,\ref{fig:urfit} shows the averaged radial velocity $\langle - U^r\rangle_{\rho, t \phi \theta}$ plotted with black dots. The Offset Geodesic model (Eq.\,\ref{eq:offsetgeo}) and the Pure Geodesic model (Eq.\,\ref{eq:puregeo}) are plotted in orange and green respectively. For the Offset model, we fix $u_I$ to the value of the first point on the simulated profile that is within the ISCO ($u_I=(0.0136\pm0.0007)c$).  As expected, this value is small and is quickly dwarfed by the $\mathcal{O}(1)c$ geodesic term as the plunging flow approaches the horizon. Both models show a remarkable agreement with the simulation, with the residuals quickly dropping to the percentage level towards the horizon. This demonstrates that in a weakly magnetised thin-disc epoch, the dominant forcing term in the plunging region relativistic Euler equation is gravity. Pressure gradients and magnetic fields are sub-dominant, which was a fundamental assumption of the MB23 model.
\par
In the right-hand panel of Fig.\,\ref{fig:urfit} we plot the averaged radial profile of the angular momentum, $\langle U_\phi \rangle_{\rho, t \phi \theta}$. We see a clear but relatively small drop ($\delta_\mathcal{J}\approx 5.3 \%$) in the angular momentum over the plunge. Interestingly, this is consistent with the continuum fitting results of \cite{mummeryContinuumEmissionPlunging2024} who parametrise the intra-ISCO thermodynamics in terms $\delta_{\cal J}$, and by fitting the X-ray spectra of  the black hole X-ray binary MAXI J1820 + 070, found $\delta_{\cal J} \approx 4\%$. The horizontal green line represents the constant angular momentum associated with a circular orbit at the ISCO, which approximates the true angular momentum to within a few percent. This drop in angular momentum is the result of the outward transport of angular momentum from the plunging fluid to the main disc by a non-zero stress. This transport is small enough so that geodesic inspirals remain an excellent approximation to the dynamics of the flow. Nonetheless, this stress still extracts free energy from the disc shear. When multiplied by the considerable shear gradient, it leads to significant heating of the fluid as it passes over the ISCO and begins to plunge. To illustrate the significance of this heating, if one solves for the disc thermodynamics within the \cite{novikovAstrophysicsBlackHoles1973} framework, including an ISCO stress equivalent to what is found in our simulation, the central temperature is hottest \emph{at the ISCO itself} (see Fig. E4 of MB23). This is in contrast to the zero-stress limit where this temperature vanishes at the ISCO. In this way, a dynamically unimportant stress in the plunging region is thermodynamically crucial.
\par
\begin{figure*}
    \centering
    \includegraphics[width=0.49\linewidth]{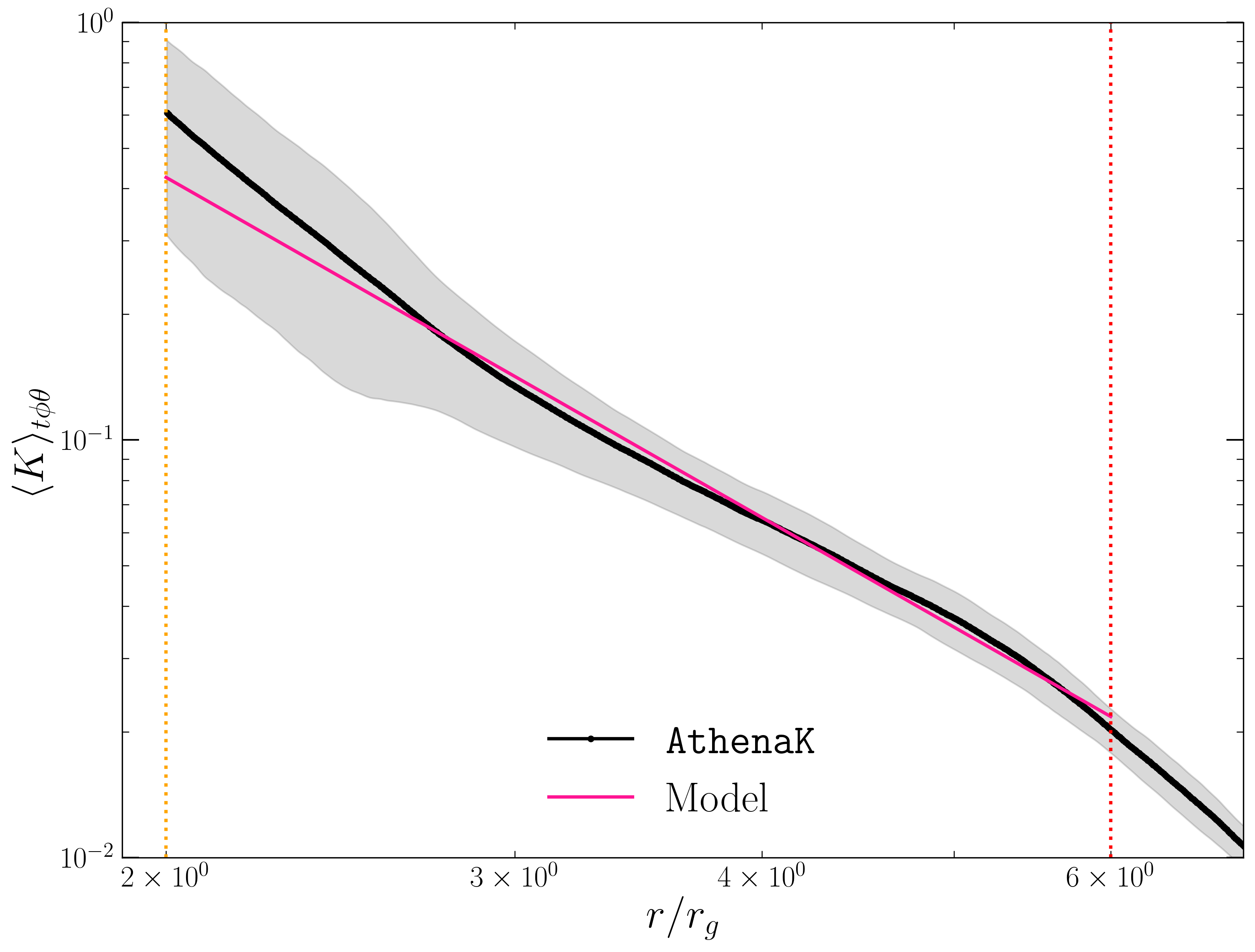}
    \includegraphics[width=0.49\linewidth]{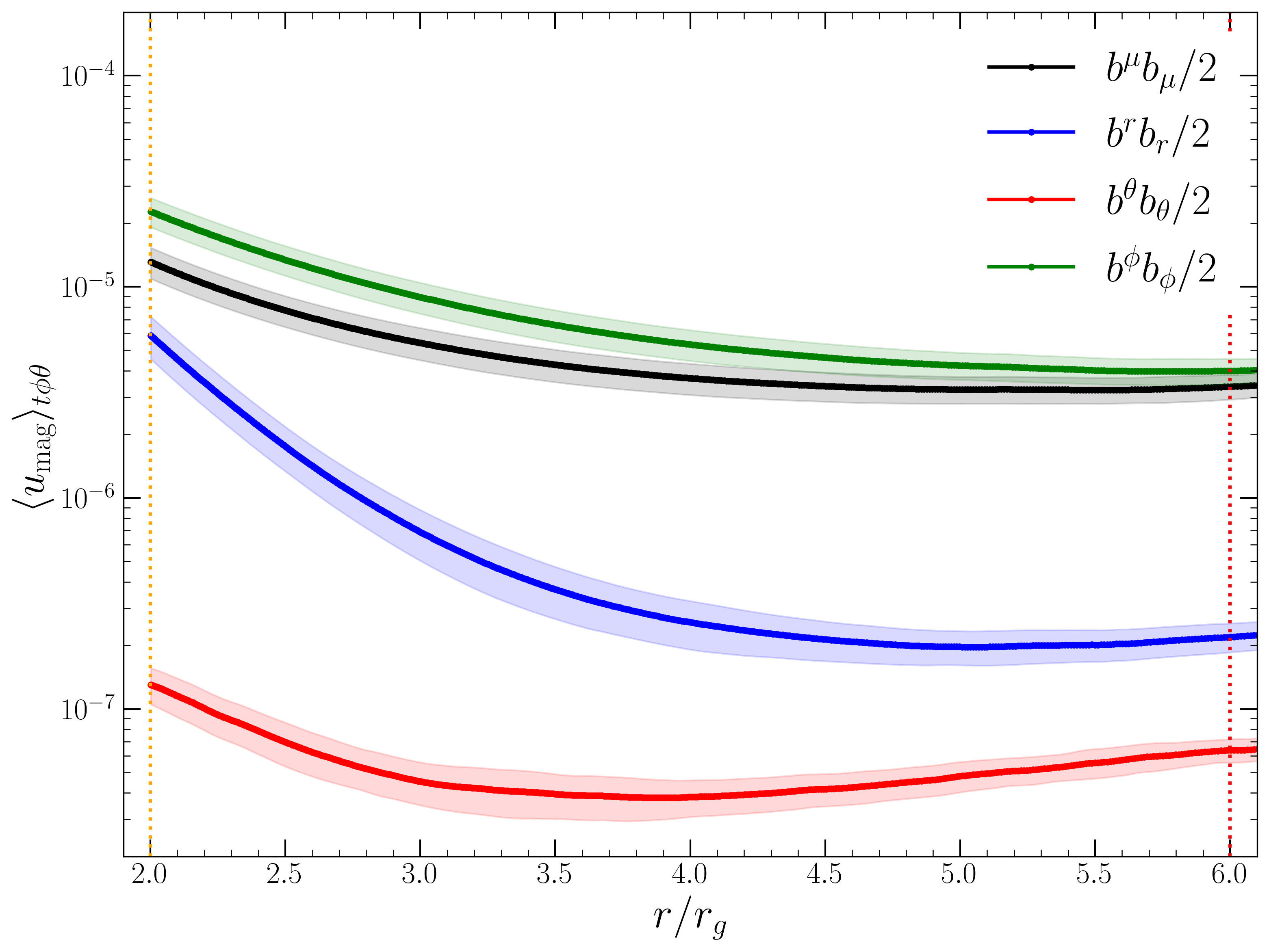}
    \includegraphics[width=0.9\linewidth]{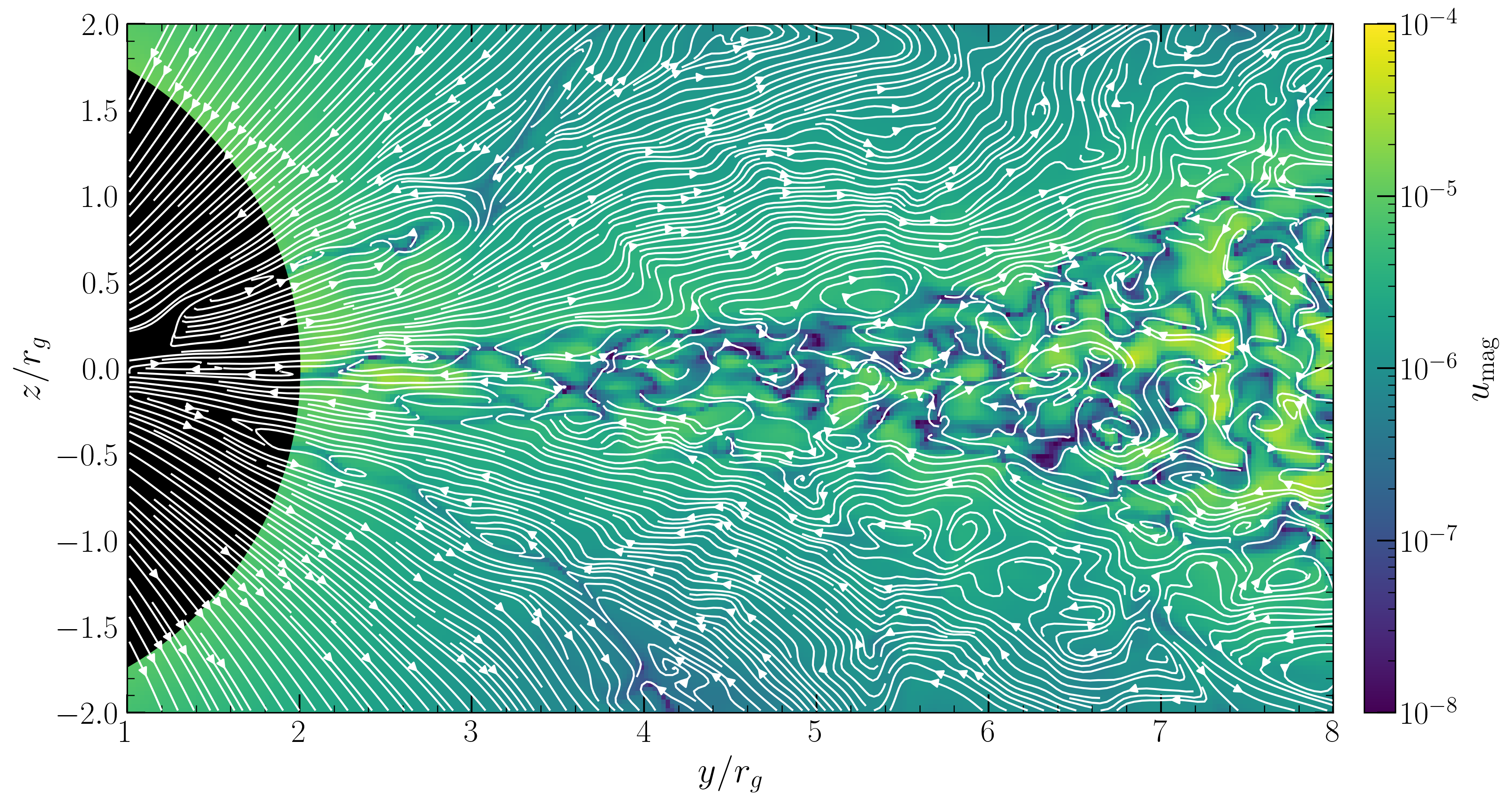}
    \caption{The plunging region specific entropy profile represented by $K=P\rho^{-\gamma}$ (upper-left), magnetic energy profiles (upper-right) and a snapshot of the $x=0$ slice taken at $t=20,000 \, t_g$ (lower). In the upper-left, black dots show the temporally ($[15,20]\text{k}t_g$), azimuthally ($[0,2\pi]$) and vertically ($[76^\circ,104^\circ]$) averaged simulated specific entropy. The shaded area represents a $\pm1\sigma$ standard deviation. In pink is a fitted power law model $K=K_I(r/r_I)^{-m}$. In the upper-right, black dots show the similarly averaged electromagnetic energy density ($u_\mathrm{mag}=b^\mu b_\mu/2$) profile, with individual components in colour (note that $b^tb_t$, which is not shown, is negative). Below, the electromagnetic energy density is shown in colour, with magnetic field lines overlaid in white. A current sheet can be seen in the mid-plane of the plunging region where oppositely directed field lines reconnect at the grid-scale.}
    \label{fig:EntropyandMagField}
\end{figure*}

The upper left-hand side of Fig.\,\ref{fig:EntropyandMagField} shows the averaged specific entropy profile, represented by $ \langle K \rangle  = \langle P\rho^{-\gamma}\rangle_{t \phi \theta}$. We notice that $\langle K \rangle$ rises throughout the plunge and so contrary to the assumptions of the MB23 model, the plunge is not perfectly adiabatic. We find that this rise can be reasonably well-modelled as a radial power law $\langle K \rangle=K_I(r/r_I)^{-m}$ where $m$ is an index and $K_I$ is a normalisation, both of which we fit for by minimising the error-weighted squared distance between the simulated data and the model. These parameters (excluding normalisations) and their associated errors are summarised in Table\,\ref{tab:parameters}.
\begin{table}
\centering
\begin{tabular}{llll}
\hline
Model  & Quantity                                   & $\epsilon$ & $m$                  \\ \hline
Eq.\,\ref{eq:offsetgeo} & $\langle-U^r\rangle_{\rho, t \phi \theta}$ & $\it{0.041\pm0.002}$      & -                     \\
Power Law     & $\langle K \rangle_{t \phi \theta}$        & -          & $2.71$            \\
Eq.\,\ref{eq:surfdensitymodel}      & $\langle \Sigma \rangle_{t \phi}$          & $0.021$      & -              \\
Eq.\,\ref{eq:densitymodel}     & $\langle \rho \rangle_{t \phi \theta}$     & $0.015$      & $\ast$ 2.71          \\
Eq.\,\ref{eq:pressuremodel}     & $\langle P \rangle_{t \phi \theta}$        & $0.015$      & $\ast$ 2.71         \\
Eq.\,\ref{eq:tempmodel}    & $\langle T \rangle_{t \phi \theta}$        & $0.044$      & $\ast$ 2.71          \\\hline
\end{tabular}
\caption{
Summary of the best-fit parameters for each model (excluding normalisations). With the exception of the radial velocity, these are found by minimising the error-weighted squared distance to the simulated data. An asterisk ($\ast$) indicates that the parameter was fixed at the given value during the fitting process. The parameter $\epsilon$ is related to the ISCO velocity (Eq.\,\ref{eq:epsilon}). For the radial velocity, $\epsilon$ is directly calculated from the first point on the simulated profile that is within the ISCO and is \emph{not} a fitted parameter. The parameter $m$ is the fitted index used to model the radial dependence $\langle K \rangle=K_I(r/r_I)^{-m}$.
}
\label{tab:parameters}
\end{table}
\par
Physically, the rise in entropy is an interesting result as, naively, it appears unlikely that the flow should be dissipating so much in a region where the flow is well described by geodesic plunges onto the black hole. We suggest that the extra heating is due to grid-scale magnetic reconnection caused by the combination of a  rapid radial acceleration and vertical compression within the plunging region, which cause field lines to be stretched (by flux freezing) towards the equatorial plane. If field lines of opposing polarity (initially) above and below the equatorial plane both undergo this same dynamical evolution and meet, they will reconnect and heat the flow. To demonstrate that this is indeed what is happening in our simulation, the lower panel of Fig.\,\ref{fig:EntropyandMagField} shows an extended current sheet in the mid-plane of the plunging region (taken from a snapshot from the latest time in our simulation), with reconnecting oppositely directed field lines either side. This heating via magnetic reconnection has been noted before in the literature \citep[e.g.,][]{pennaSimulationsMagnetizedDiscs2010}.
In the upper-right panel of Fig.\,\ref{fig:EntropyandMagField}, we also plot the electromagnetic energy density profile $u_\mathrm{mag}=b^\mu b_\mu/2$ and its spatial constituents. Here, $b^\mu$ are the covariant components of the magnetic field 4-vector (e.g. \cite{gammieHARMNumericalScheme2003}). Flux freezing is clearly demonstrated by the sharp rise in the radial component (in blue) as the flow approaches the horizon. It is also clear that the magnetic field is predominantly toroidal at this stage of the simulation.
\par
\begin{figure*}
    \centering
    \includegraphics[width=0.49\linewidth]{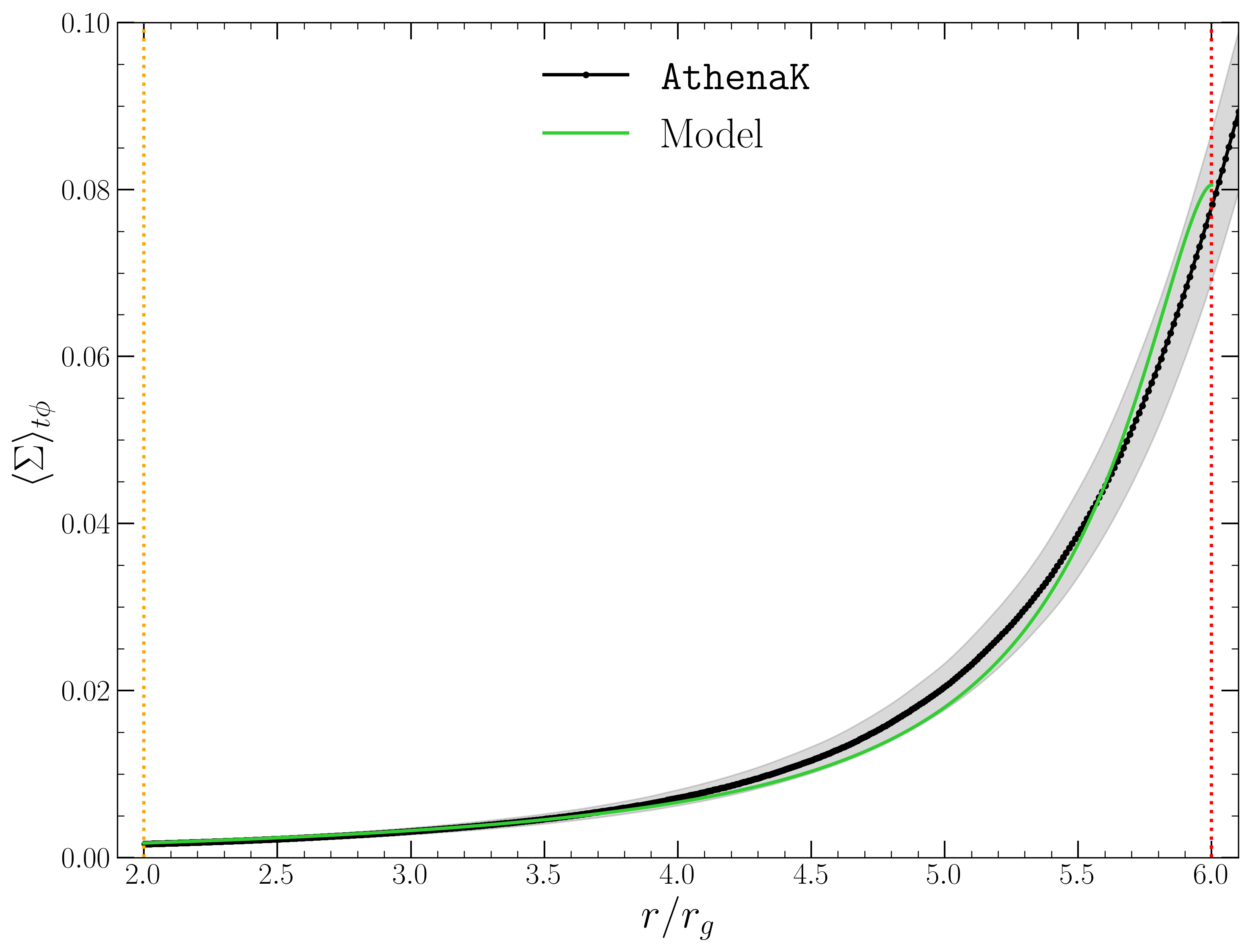}
    \includegraphics[width=0.49\linewidth]{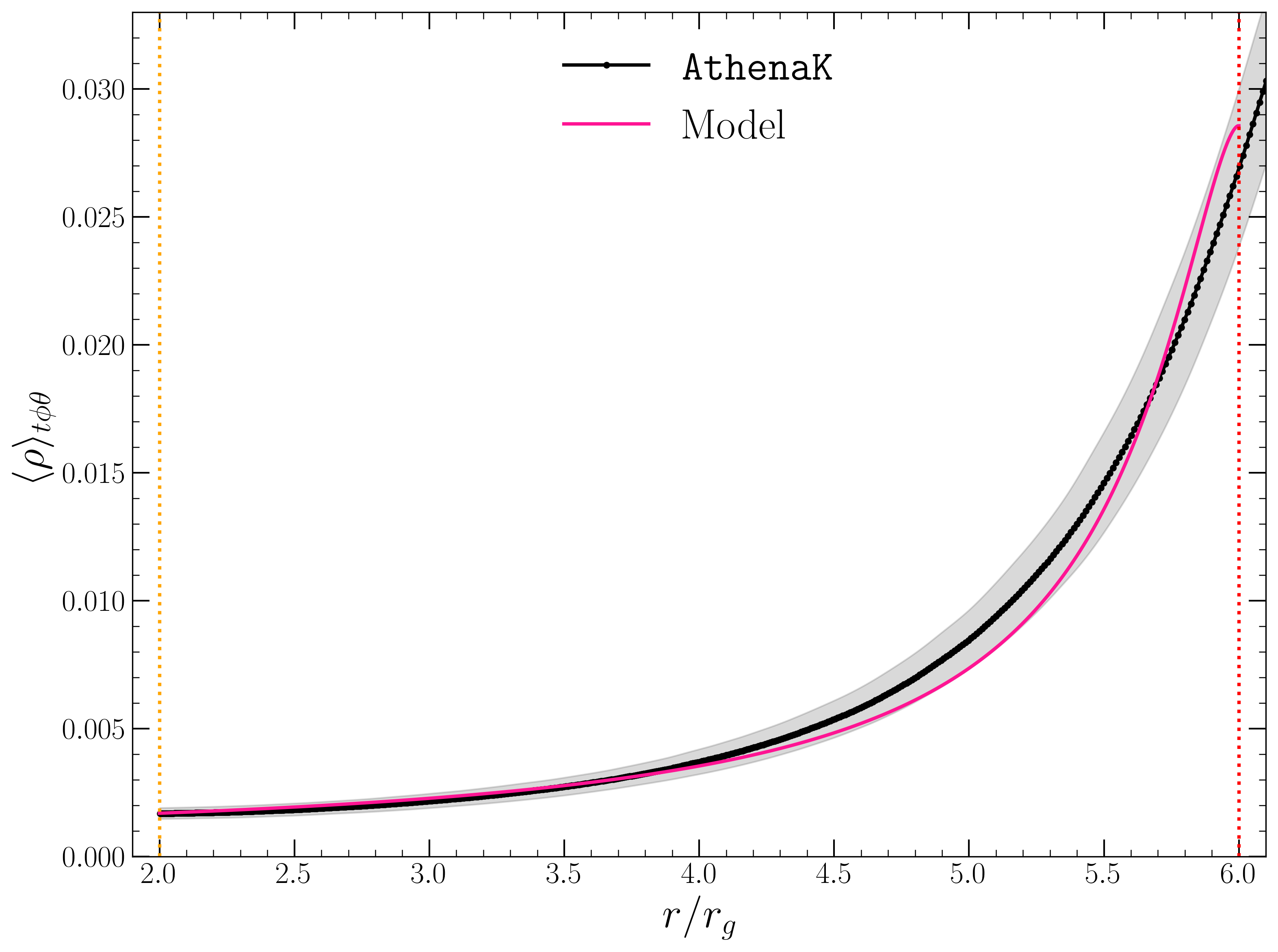}
    \includegraphics[width=0.49\linewidth]{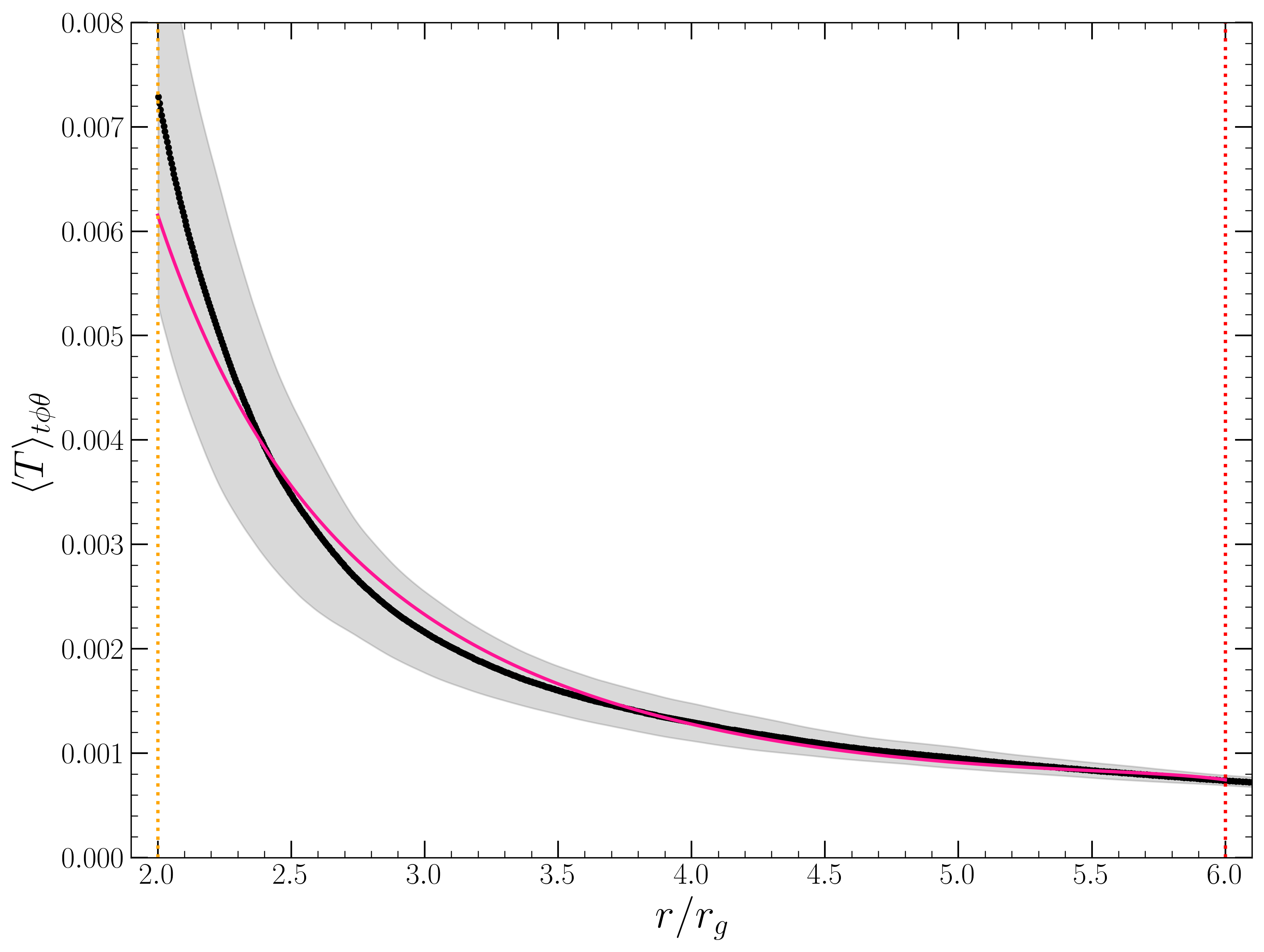}
    \includegraphics[width=0.49\linewidth]{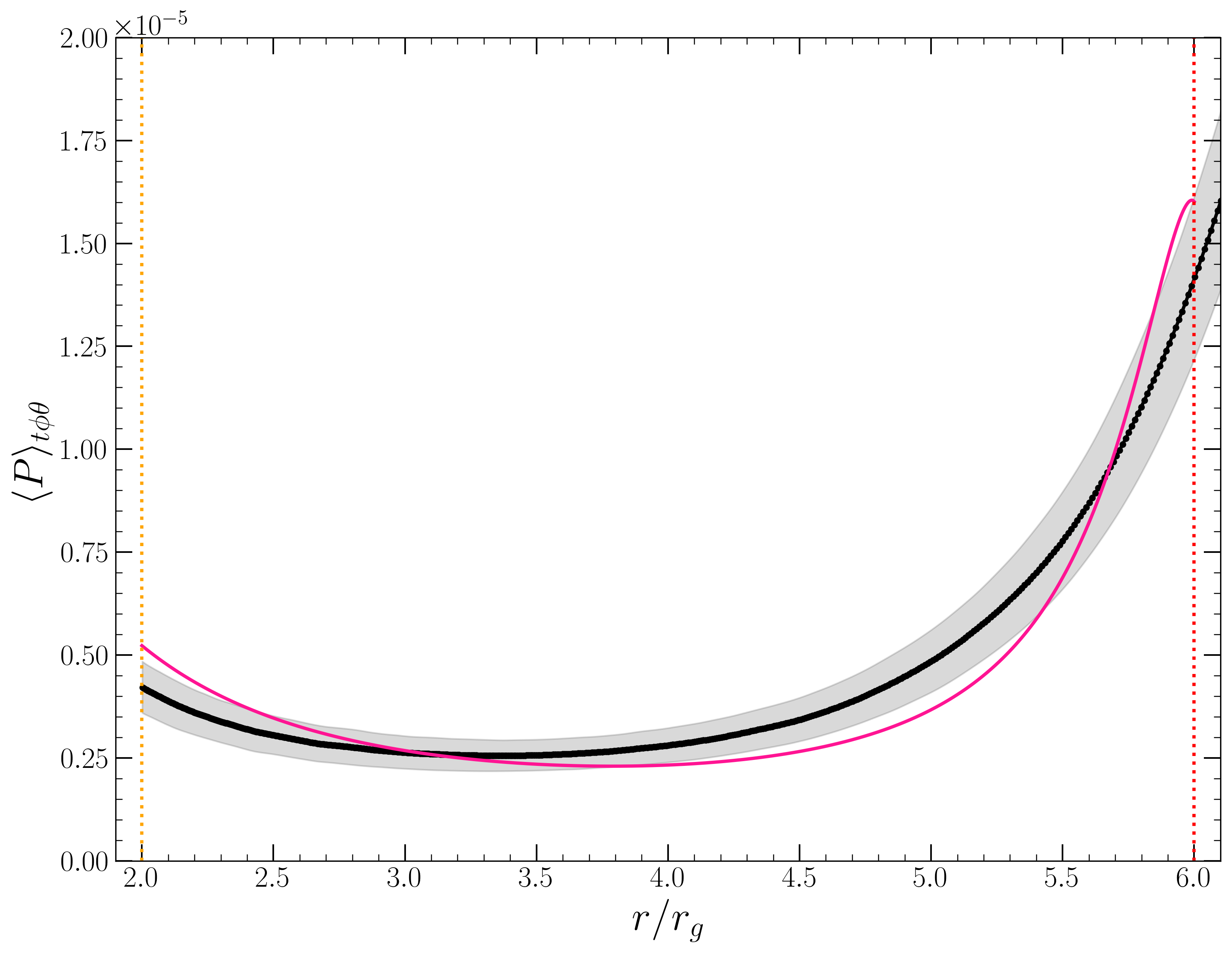}
    \caption{
    The plunging region surface density (top-left), density (top-right), pressure (bottom-right) and temperature (bottom-left) profiles. Black dots show the temporally ($[15,20]\text{k}t_g$), azimuthally ($[0,2\pi]$) and (for the density, pressure and temperature) vertically ($[76^\circ,104^\circ]$) averaged simulated quantities. The shaded regions represent $\pm1\sigma$ standard deviations. In the top-left panel, the green line is the MB23 surface density model (Eq.\,\ref{eq:surfdensitymodel}), which follows directly from mass conservation. In the other panels, the pink lines are fitted MB23 models (Eqs.\,\ref{eq:densitymodel}, \ref{eq:pressuremodel} and  \ref{eq:tempmodel}). For these fits, $K$ has been modelled with the power law model (pink) shown in Fig.\,\ref{fig:EntropyandMagField}. The density, pressure and temperature profiles are fitted separately so that they each have their own $\epsilon$ parameter (see Table\,\ref{tab:parameters}). The index $m$ is fixed to the value found from the $K$ profile.
    }
    \label{fig:thermofits}
\end{figure*}

The surface density, density, pressure and temperature profiles are plotted in Fig.\,\ref{fig:thermofits}. Using the power-law model to describe the radial dependence of $K$, we plot the MB23 models for each quantity, fitting $\epsilon$ separately for each profile by minimising the error-weighted squared distance to the simulated data. 
The best-fit $\epsilon$ parameters that we find for each model and quantity are summarised in Table\,\ref{tab:parameters}.
\par
We find excellent agreement between the MB23 model for the surface density (Eq.\,\ref{eq:surfdensitymodel}) and the simulated data. This is unsurprising. We have already demonstrated excellent agreement between the simulated radial velocity and the offset geodesic model (Eq.\,\ref{eq:offsetgeo}), so it is clear that the surface density should also follow directly from mass conservation (Eq.\,\ref{eq:massconsv}), provided that the mass accretion rate ($\dot{M}$) is constant. We continue to find good agreement between the MB23 models for the other thermodynamic quantities ($\rho$, $P$ and $T$) and the simulated data, provided that we adequately model the radial dependence of $K$. This does not follow from mass conservation alone, and demonstrates that the MB23 scale height prescription is also robust. Note that we expect some level of disagreement for these quantities due to the fact that our model for $K$ is ad-hoc.
\par
Examining the best-fit $\epsilon$ parameters found for each quantity in Table\,\ref{tab:parameters}, we find close agreement between the density and pressure, but some spread when one includes the temperature and surface density (although these still agree at the factor 3 level). This spread is to be expected since we are fitting functions which each have a non-linear dependence on $\epsilon$ to averaged data. For non-linear functions, the average of the function is not equivalent to the function of the average, so the best-fit $\epsilon$ is \emph{not} a measurement of the average radial velocity at the ISCO, but is likely to be broadly similar in magnitude. More technically, this is because the first derivative of the objective function ($\partial\chi^2/\partial\epsilon$), which is formally equal to zero for the best-fit $\epsilon$, will also be highly non-linear. The value of $\epsilon$ should instead be treated as an effective parameter of the MB23 model. 
\par
\begin{figure*}
    \centering
    \includegraphics[width=0.49\linewidth]{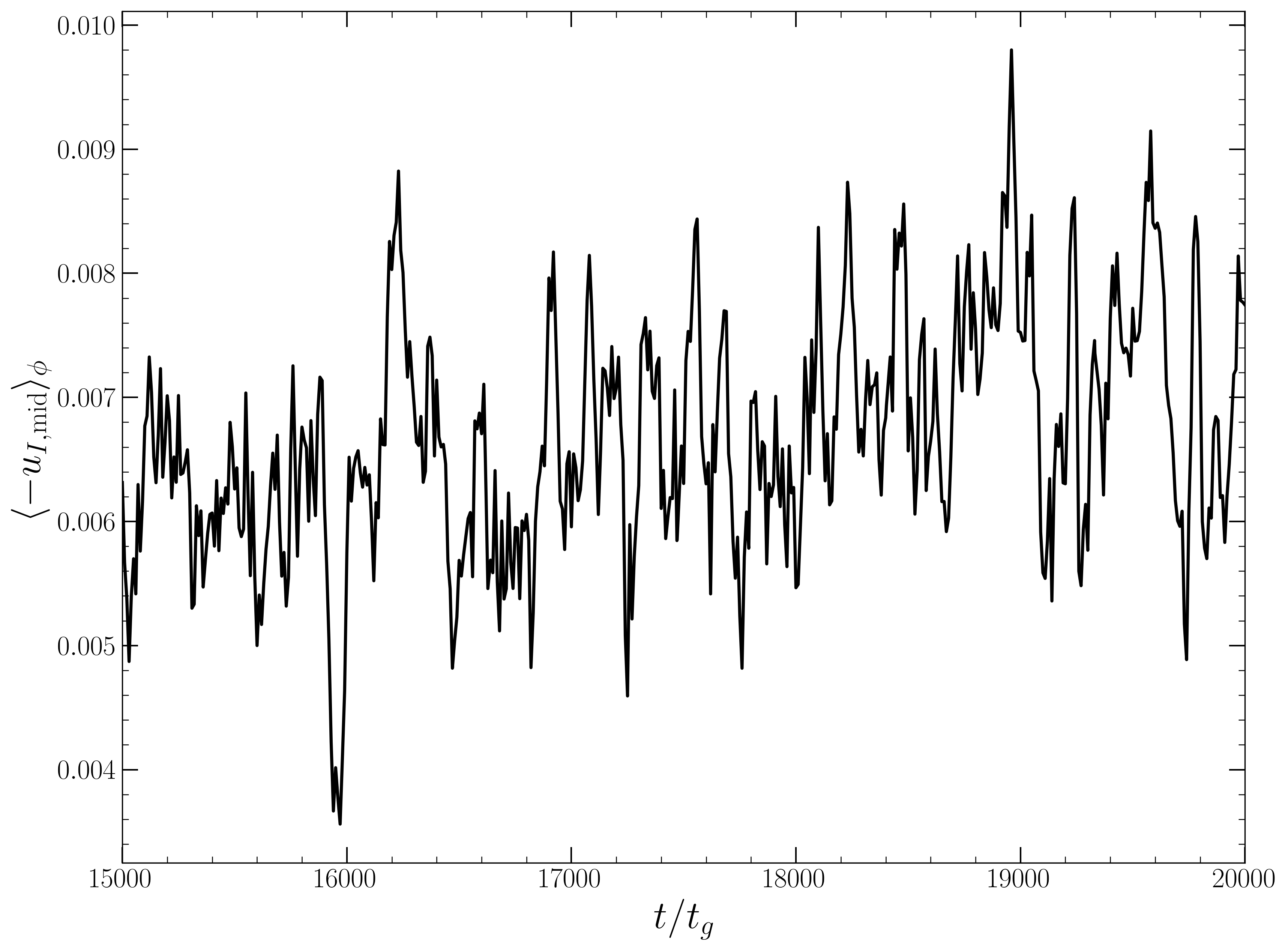}
    \includegraphics[width=0.49\linewidth]{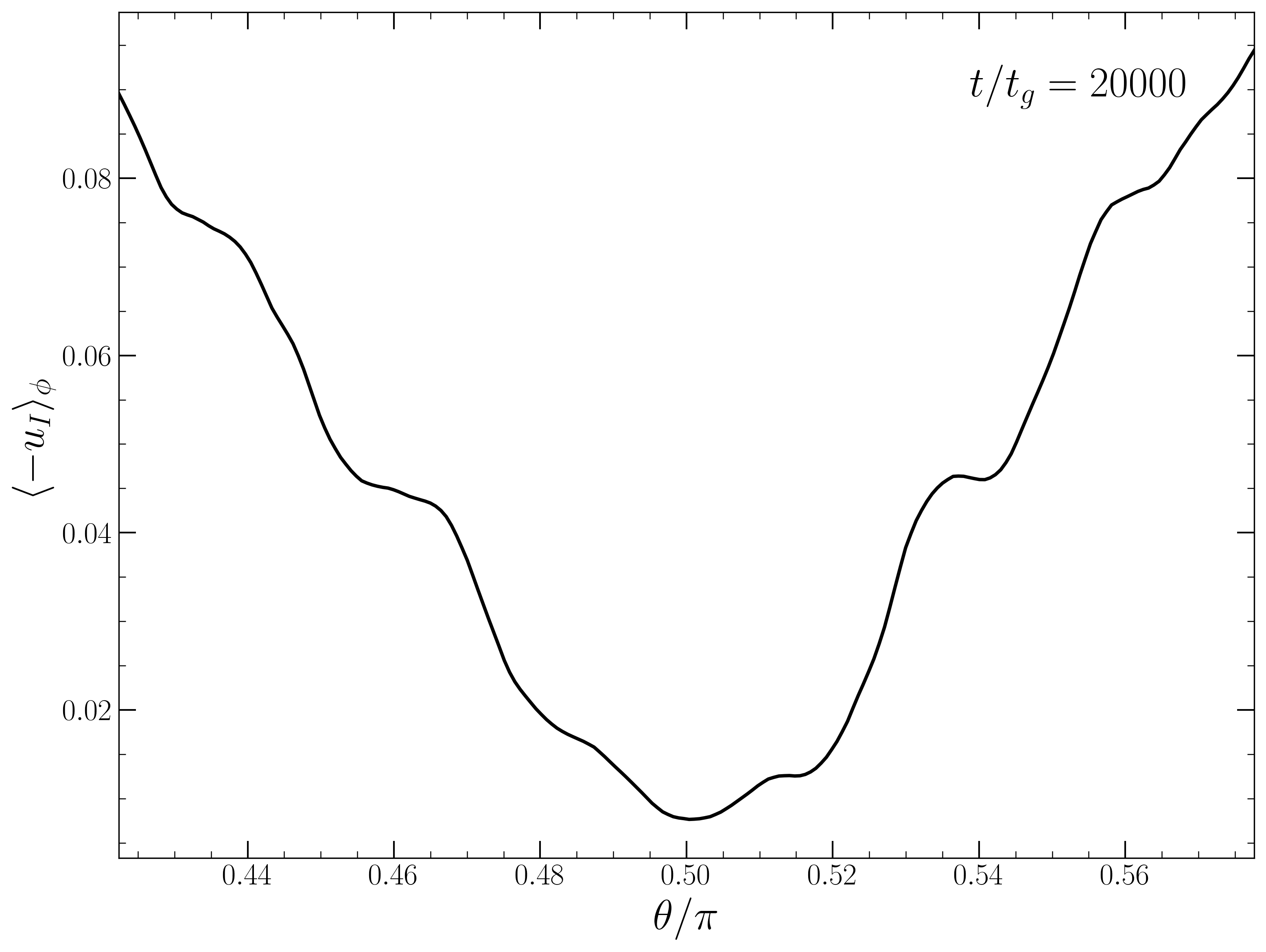}
    \includegraphics[width=0.49\linewidth]{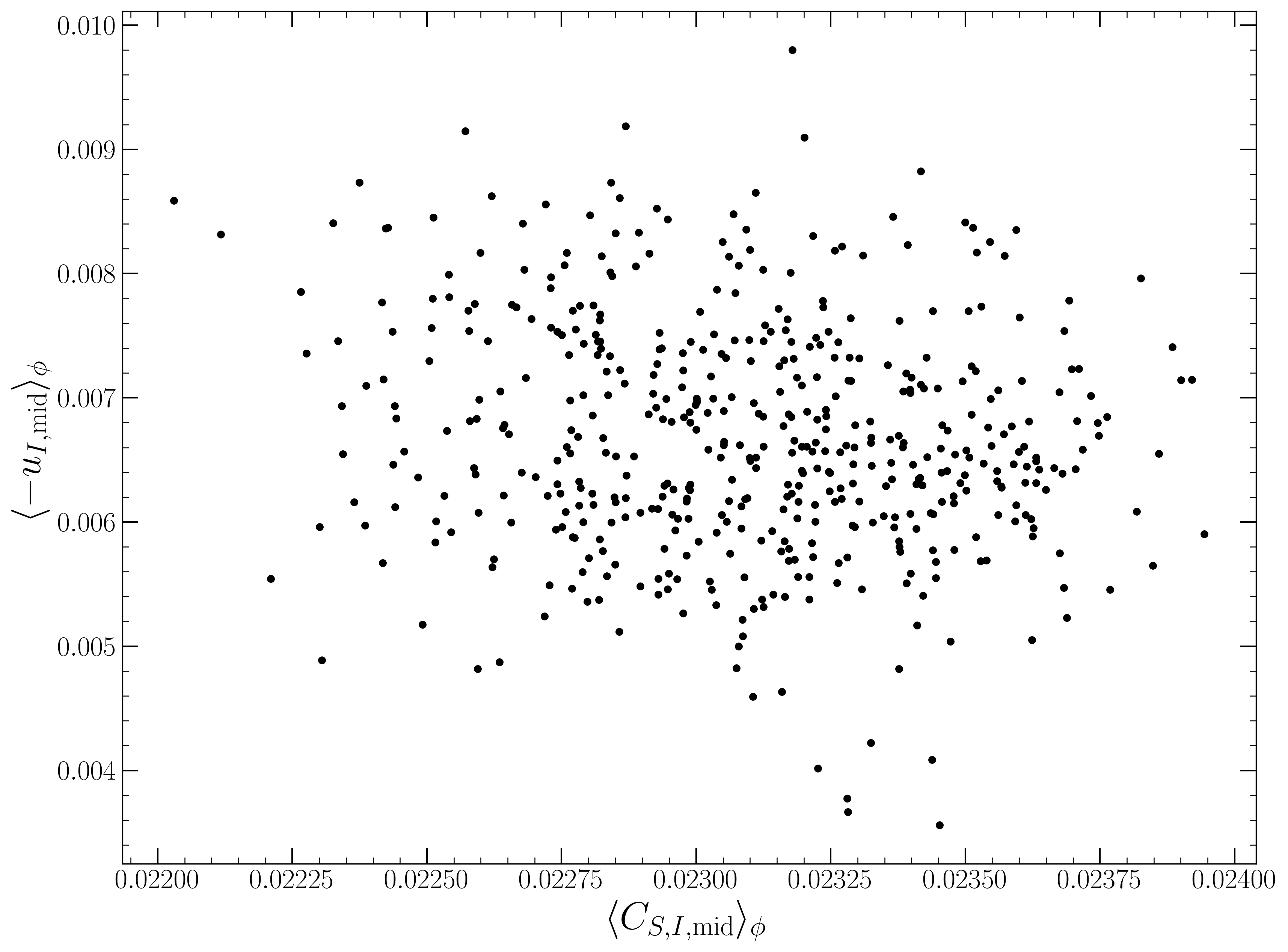}
    \includegraphics[width=0.49\linewidth]{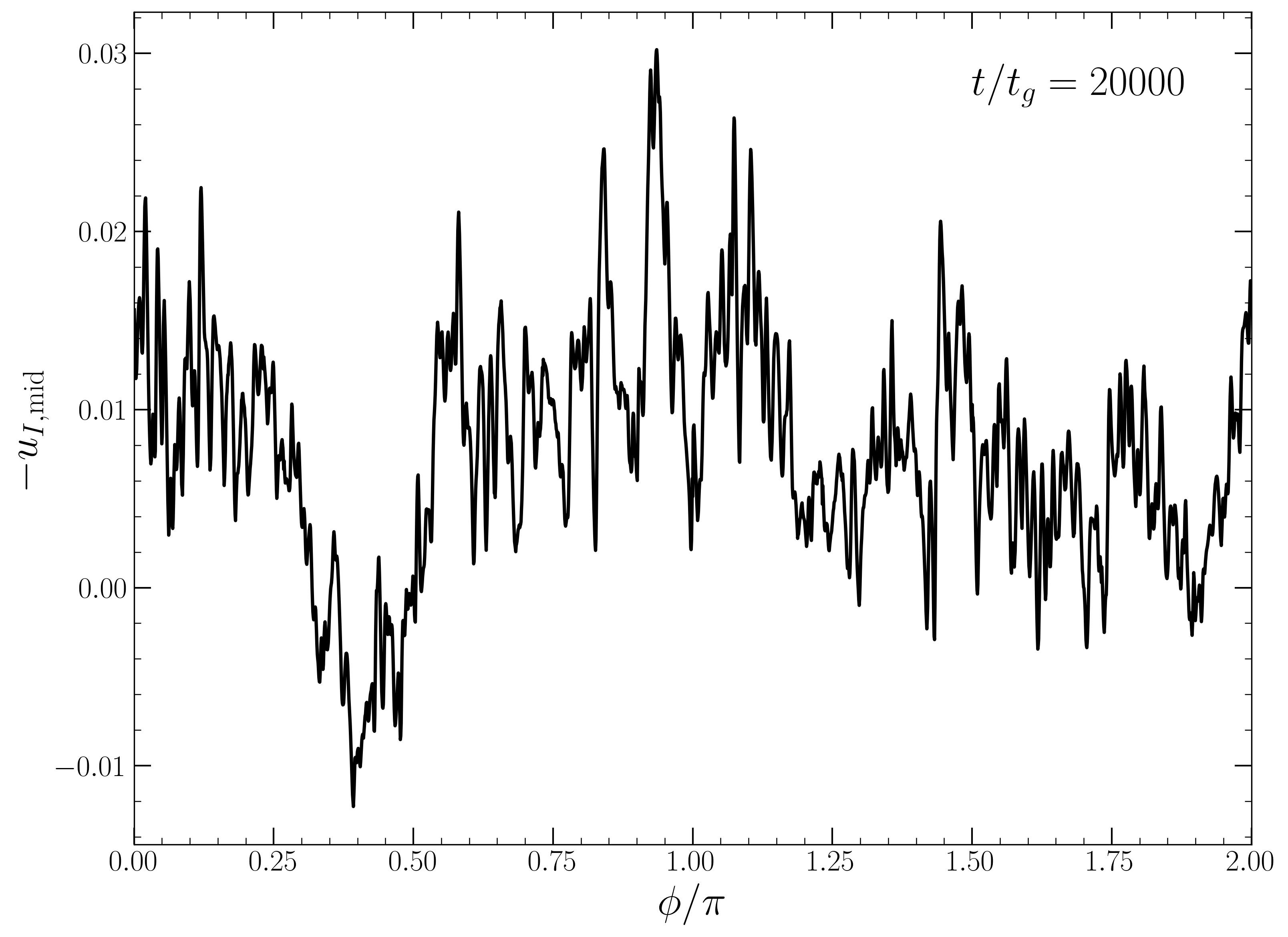}
    \caption{The inward radial velocity at the ISCO ($-u_I$) as a function of time (top-left), vertical height ($\theta$, top-right), azimuthal position ($\phi$, bottom-right) and azimuthally averaged mid-plane sound speed at the ISCO ($\langle C_{S,I,\text{mid} \rangle_\phi}$, bottom-left). To marginalise over other variables, the averaging procudure is slightly different in each case. For the temporal variation, we plot the azimuthally averaged mid-plane ISCO velocity at each time. For the vertical variation, we plot the azimuthally averaged ISCO velocity at the time $t/t_g = 20,000$. For the azimuthal variation, we plot the mid-plane value at the time $t/t_g = 20,000$. Finally, for the sound speed we plot the same quantity as for the temporal variation. Each point is measured at a common time. 
    }
    \label{fig:UIVariation}
\end{figure*}

Fig.\,\ref{fig:UIVariation} shows the simulated radial velocity at the ISCO ($u_I$) as a function of time, height, azimuth and the ISCO sound speed. It rapidly fluctuates as a function of azimuth and time, highlighting that the flow is still very much turbulent as it passes over the ISCO and is yet to transition to the more laminar plunging flow \citep[e.g.][]{mummeryDynamicsAccretionFlows2024}. In the lower-left panel, we find no correlation between the azimuthally averaged mid-plane isothermal sound speed and the azimuthally averaged ISCO velocity. A curious observation is that the mid-plane ISCO velocity is outward flowing over a fraction of the azimuth for the snapshot shown in the bottom-right. Examining the upper-left panel, we also notice that the inward ISCO velocity is at a minimum in the mid-plane and sharply rises with scale height, either side of the mid-plane.
\par
One can interpret these observations as a series of layers `sliding' over each-other in the disc atmosphere and into the plunging flow. Layers higher into the disc atmosphere experience less obstruction from turbulent eddies and so slide in more quickly. Where fast flowing layers of disc atmosphere meet at the mid-plane, they can be occasionally pushed back into the main body of the disc, explaining why there is sometimes observed \emph{outflow} in mid-plane. Since all of the simulated profiles in Fig.\,\ref{fig:thermofits} are vertically averaged, $\epsilon$ is best thought of as an effective parameter that describes the entire vertical structure. Nonetheless, we expect that the averages will be skewed towards the layers where the quantity is greatest, i.e. the mid-plane in the case of the density. Due to the vertical stratification in the flow speed, the best-fit $\epsilon$ should preferentially reflect these, for example, over-dense layers.

\begin{figure}
    \centering
    \includegraphics[width=\linewidth]{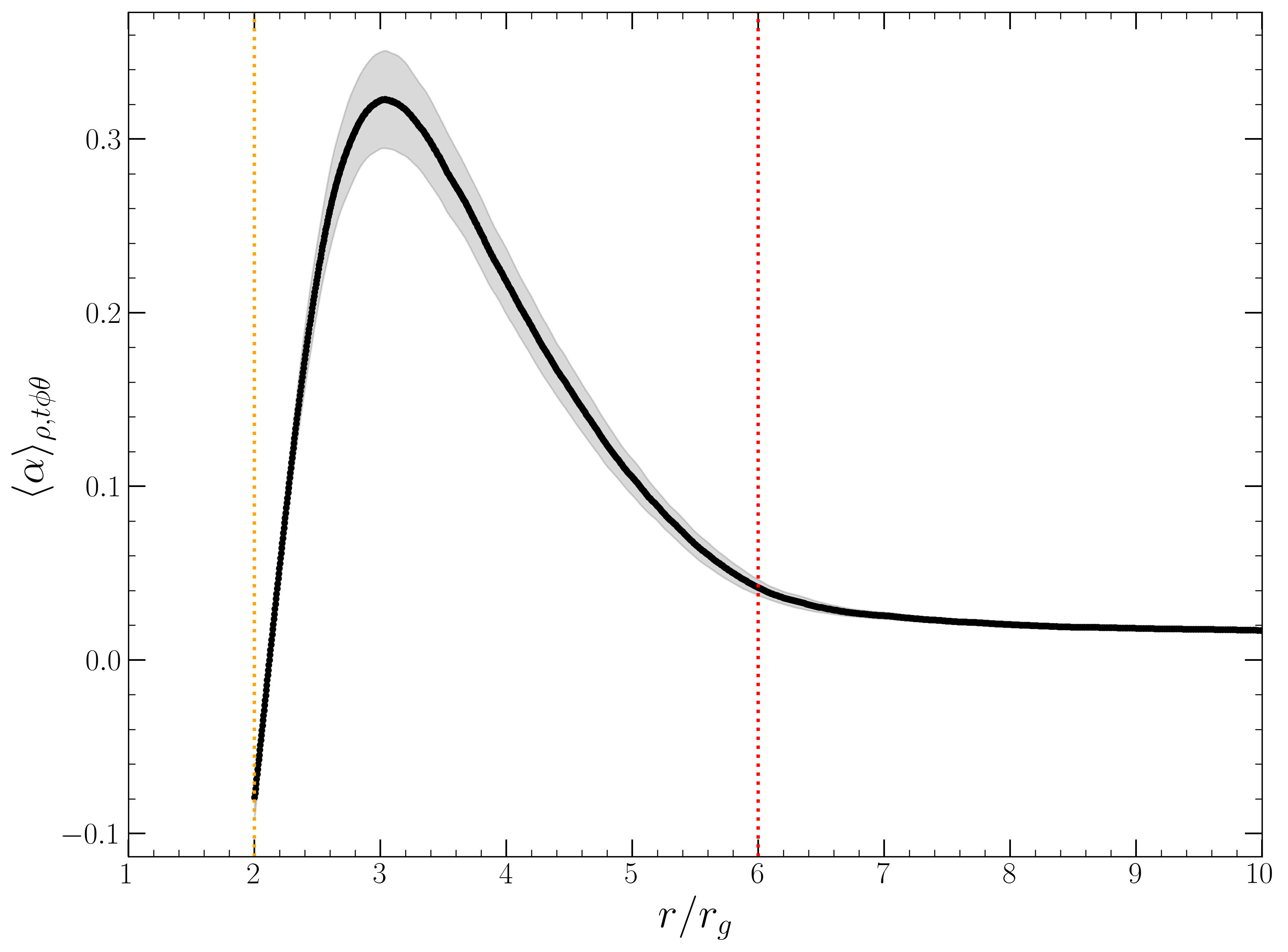}
    \caption{The plunging region $\alpha$ profile. Black dots show the density weighted temporally ($[15,20]\text{k}t_g$), azimuthally ($[0,2\pi]$) and vertically ($[76^\circ,104^\circ]$) averaged simulated $\alpha$. A clear order of magnitude rise is observed.
    }
    \label{fig:alpha_plot}
\end{figure}

In Fig.\,\ref{fig:alpha_plot}, we plot the \cite{shakuraBlackHolesBinary1973} angular momentum transport parameter, $\left\langle\alpha\right\rangle_{\rho, t \phi \theta}$ as a function of radius. We define this in the standard manner
\begin{equation}
    \alpha \equiv \frac{T^\text{mag}_{\hat{\phi}\hat{r}}}{p+b^2/2},
\end{equation}
where $T^\text{mag}_{\hat{\phi}\hat{r}}$ is the $r,\phi$ component of the fluctuating, magnetic part of the stress-energy tensor calculated in the local fluid rest frame, 
\begin{equation}
    T^\text{mag}_{\hat{\phi}\hat{r}} = e_{\hat{r}}^{\mu}e_{\hat{\phi}}^{\nu} T^\text{mag}_{\mu\nu} = e_{\hat{r}}^{\mu}e_{\hat{\phi}}^{\nu}(b^2U_\mu U_\nu + \frac{1}{2}b^2 g_{\mu\nu} - b_\mu b_\nu) . %
\end{equation}
In these expressions $e_{\hat{\alpha}}^{\mu}$ are the orthonormal basis vectors \citep{Bardeen72} of the local fluid rest frame (e.g. \cite{kulkarniMeasuringBlackHole2011}). Note that the first two terms vanish in the local rest frame of the flow.  We clearly observe that this is not a constant parameter in the plunging region. This is not surprising, as the magnetic fields are dragged by the relativistic flow into an ever decreasing region, amplifying their turbulent stress \citep[this is just flux freezing once again,][]{krolikMagnetizedAccretionMarginally1999}.
\par
This flux freezing, which is a \emph{fundamentally MHD} effect, can of course not be captured by a phenomenological viscous model. It is inevitable that within the plunging region the strength of the magnetic fields will always be amplified, owing to the gravitational focussing of the flow and Alfv\'ens theorem. A growing MHD stress can then transport a small amount of angular momentum  back to the main body of the disc, and have associated with it a large dissipation owing to the large shear gradient. This stress within the plunging region is small enough that the flow is well approximated by geodesic dynamics, but not so small that the transport (and the dissipation) is effectively zero. Effectively zero transport is a prediction of the naive $\alpha$ framework, which is clearly not valid in this regime. 

A second, possibly less consequential but still important, shortcoming of the $\alpha$ framework is its assumption of locality. In the main body of the disc the mean radial  velocity is an asymptotic order smaller than the turbulent fluctuations in the velocities, which are themselves an asymptotic order smaller than the azimuthal orbital velocity. As a consequence of this scaling, the free energy, which is extracted from the shear flow by the stress, is dissipated \emph{locally}. By stark contrast, in the plunging region, the inward radial velocity is of comparable magnitude to the azimuthal velocity and so no equivalent asymptotic scaling holds. The free energy that is extracted from the shear need not be locally dissipated, and is in fact transported back to the main body of the disc where it is dissipated in the near-ISCO region. It is therefore clear that an $\alpha$-style coupling between the \emph{local} stress and \emph{local} pressure has no meaning within the plunging region. As such, models for intra-ISCO flows based upon the $\alpha$-disc framework are physically inappropriate, and have little physical content.
\section{Conclusions}
\label{sec:Conclusion}
In this paper, we have presented a thin, weakly magnetised global simulation of the inner accretion disc around a Schwarzchild black hole using the GRMHD code {\tt AthenaK}. We have demonstrated a remarkable agreement between the MB23 analytic geodesic model of the plunging region thermodynamics and the simulated quantities. One should remember that the simulation was 3D and fully turbulent, so extracting simple self-similar solutions from its thermodynamic variables is not a trivial result. This extends the work of \cite{mummeryThreedimensionalStructureBlack2024}, who did a similar study for thick discs.
\par
We measure a small but significant ($\delta_\mathcal{J}\approx 5.3 \%$) drop in the angular momentum of the plunging fluid as it spirals inwards. This drop in angular momentum from the ISCO to the horizon barely affects the geodesic infall, but as a contributing factor to the energy dissipation rate, the resulting angular momentum flux is indicative of significant heating near to the ISCO. This ensures that the fluid retains non-vanishing thermodynamic quantities as it plunges towards the black hole. We have also measured the \cite{shakuraBlackHolesBinary1973} angular momentum transport parameter, $\alpha$, in the local rest frame of the fluid. We find an order of magnitude rise over the plunge. It is clear from this alone that a constant $\alpha$-disc framework is irrelevant in this region. Furthermore, we have argued the plunging region represents a distinct regime where the assumptions of locality underpinning the $\alpha$-disc model clearly no longer hold.
\par
Analytically modelling the plunging region thermodynamics has great potential utility for observers, particularly for calculating the emission which originates from a full black hole accretion disc that has not been artificially truncated at the ISCO. Such a calculation has already been used to explain the high energy tails of thermal X-ray binary spectra (\cite{mummeryContinuumEmissionPlunging2024}, \cite{mummeryPlungingRegionEmission2024}). Such full disc models should better inform our estimates of black hole spins and help us to better probe the strong gravity regime.
\par
It would be of great interest to explore the accuracy of these emission models by post-processing radiative transport for our existing GRMHD simulations, following the example of \cite{Noble11} and \cite{zhuEyeStormLight2012}. Moreover, it is now possible within {\tt AthenaK} to self-consistently evolve a directional grey radiation field alongside GRMHD (\cite{whiteExtensionAthenaCode2023}, \cite{stoneAthenaKPerformancePortableVersion2024}). This would allow for a careful examination of the importance of the radiation field on the dynamics of the plunging fluid and is an exciting avenue for further study.
\par
Finally, it is clear that there is a large parameter space to explore beyond the Schwarzchild black hole simulation that we have presented in this paper. We plan to present further simulations, with different black hole spins, initial magnetic field geometries and cooling prescriptions in the near future.

\section*{Acknowledgements}
AM was supported by a Leverhulme Trust International Professorship grant [number LIP-202-014]. The Center for Computational Astrophysics at the Flatiron Institute is supported by the Simons Foundation. An award for computer time was provided by the U.S. Department of Energy’s (DOE) Innovative and Novel Computational Impact on Theory and Experiment (INCITE) Program. This research used resources from the Argonne Leadership Computing Facility, a U.S. DOE Office of Science user facility at Argonne National Laboratory, which is supported by the Office of Science of the U.S. DOE under Contract No. DE-AC02-06CH11357. The authors would like to acknowledge the use of the University of Oxford Advanced Research Computing (ARC) facility in carrying out this work. http://dx.doi.org/10.5281/zenodo.22558

\section*{Data Availability}

No observational data was used in producing this manuscript. Numerical results will be shared upon reasonable request with the corresponding author.



\bibliographystyle{mnras}
\bibliography{andy,jake} 




\appendix

\section{Lower Resolution Simulation}\label{appA}
As a proof of concept and to develop our analysis pipelines, we also ran a lower resolution simulation with almost the same setup. We present the results of this simulation here for completeness. Table\,\ref{tab:lowres_refinement} describes the static mesh refinement configuration for this simulation. Of particular note, this simulation has an extra refinement level and double the minimum cell
spacing when compared to the high resolution simulation presented in the main body of this paper.
\par
The other notable difference is that in this simulation we experimented with an additional backup for computing the local thermal energy when the standard inversion routine led to inaccurate values (see \S \ref{sec:Simulation}). This was based on evolving the conserved entropy (i.e. $\nabla_\alpha(\rho K U^\alpha)=0$) alongside the other GRMHD conservation laws and using the independently evolved value of $K$ to compute the thermal energy for inaccurate cells. Unfortunately, this conservation law only holds for ideal gases and is not valid near to shock fronts or close to other non-ideal environments, such as reconnecting current sheets as we find in the plunging region. We found that the physical regimes where this backup was valid were too restrictive and rendered it virtually useless, so we discarded this approach for the high-resolution case to save on computational cost.
\par
\begin{figure}
    \centering
    \includegraphics[width=\linewidth]{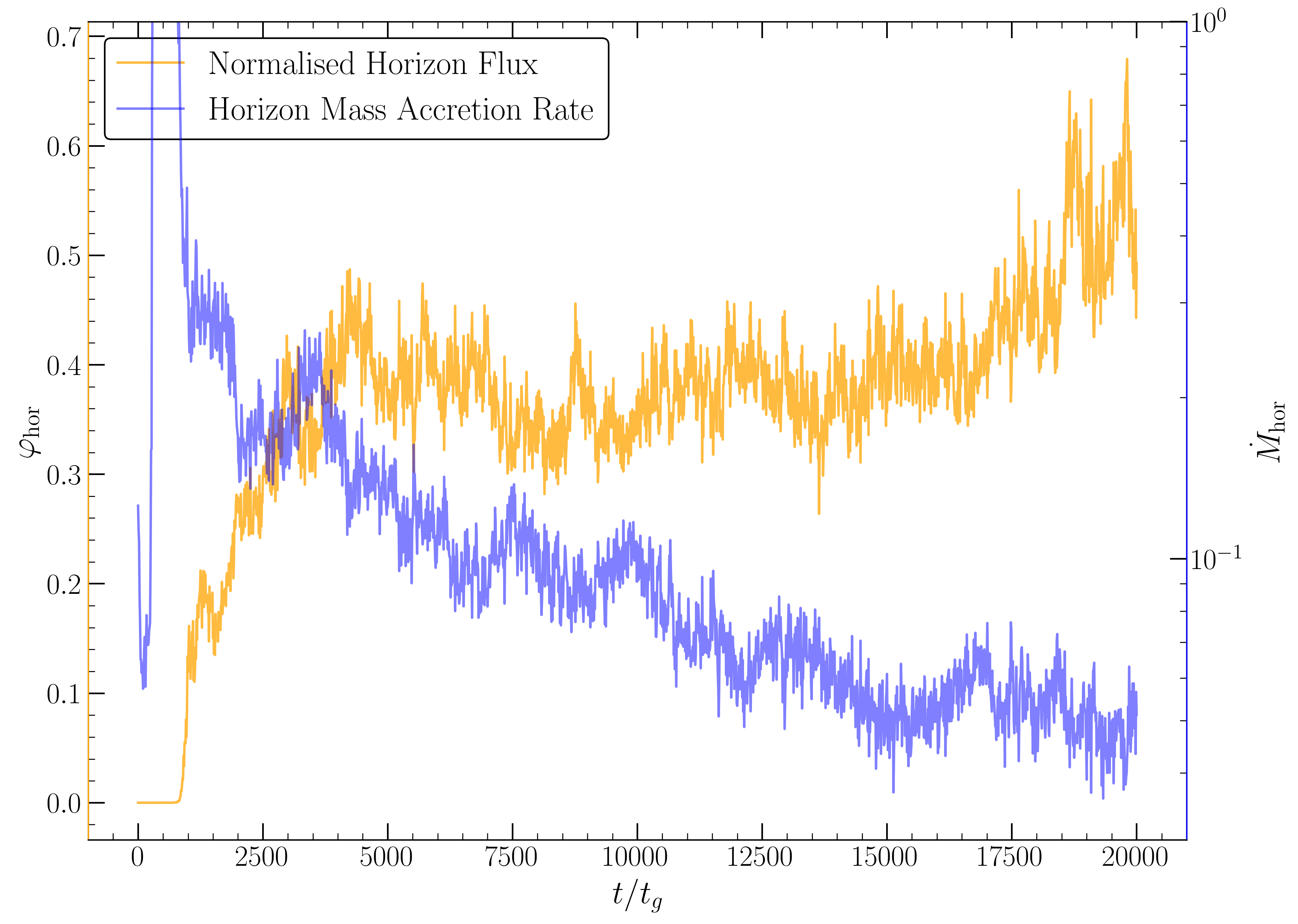}
    \caption{
    Identical to Fig.\,\ref{fig:MdotandPhi} for the lower resolution simulation. The time evolution of the mass accretion rate and the normalised magnetic horizon flux 
    }
    \label{fig:lowres_MdotandPhi}
\end{figure}
\begin{figure*}
    \centering
    \includegraphics[width=0.49\linewidth]
    {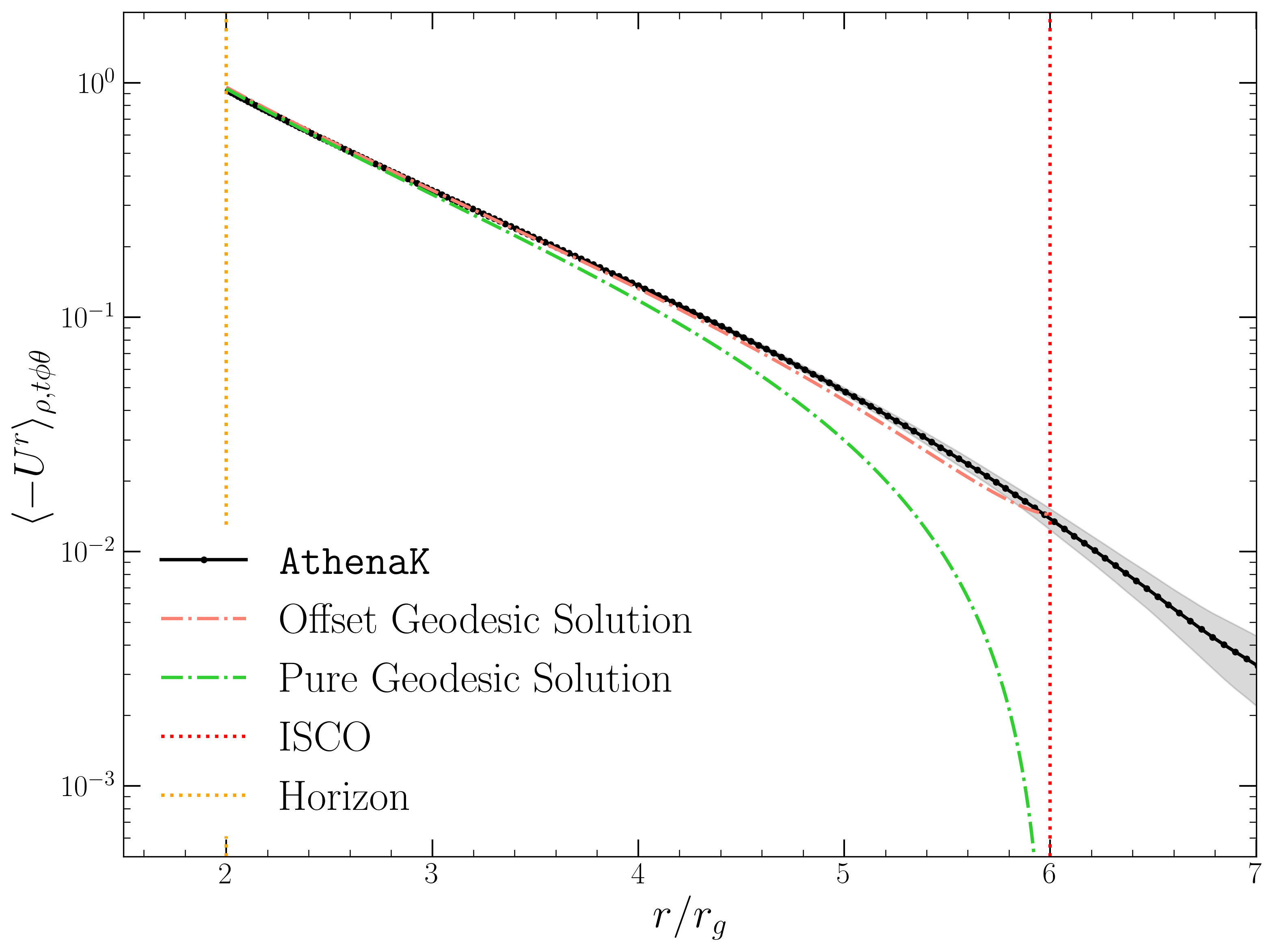}
    \includegraphics[width=0.49\linewidth]
    {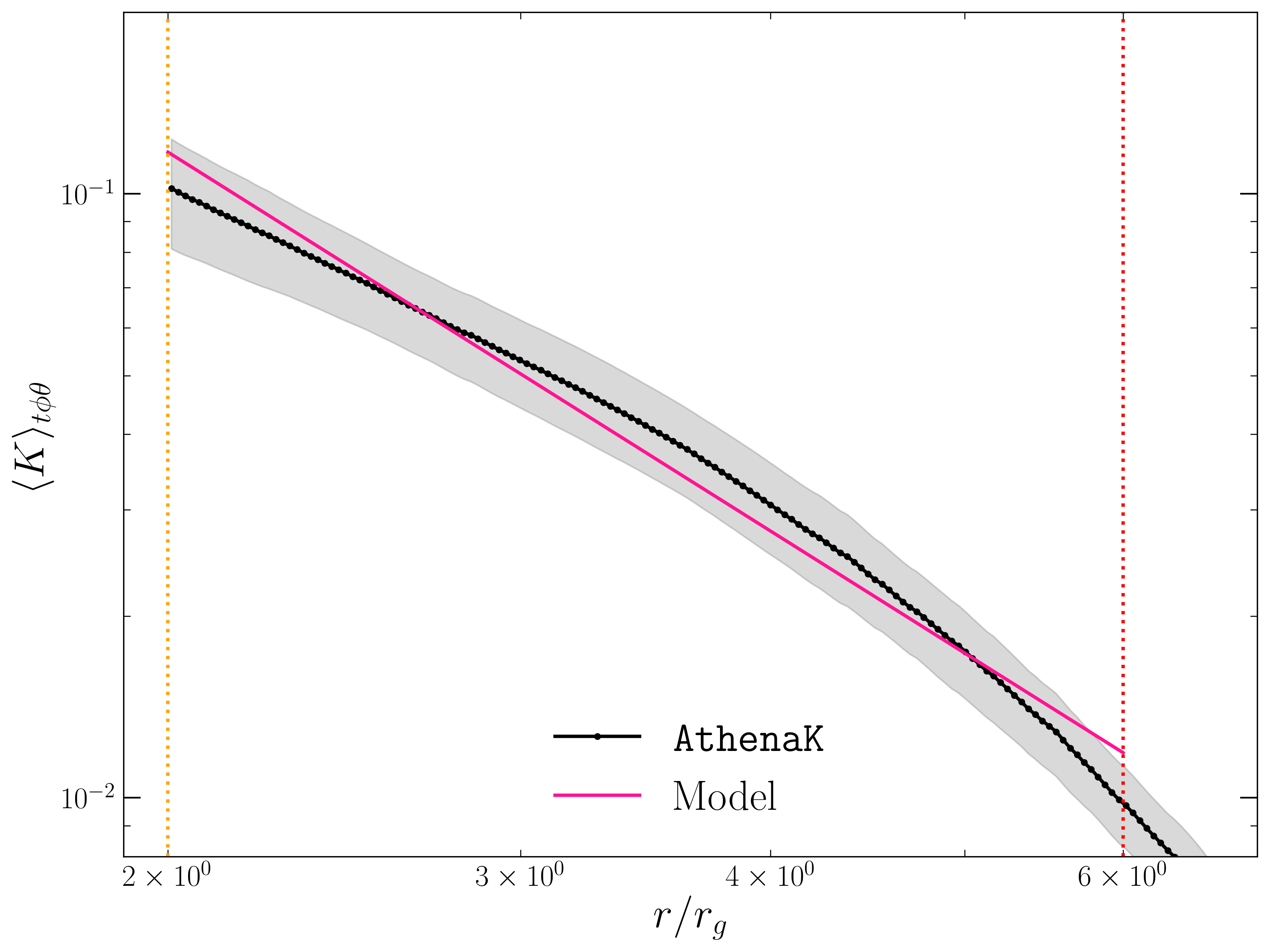}
    \includegraphics[width=0.49\linewidth]{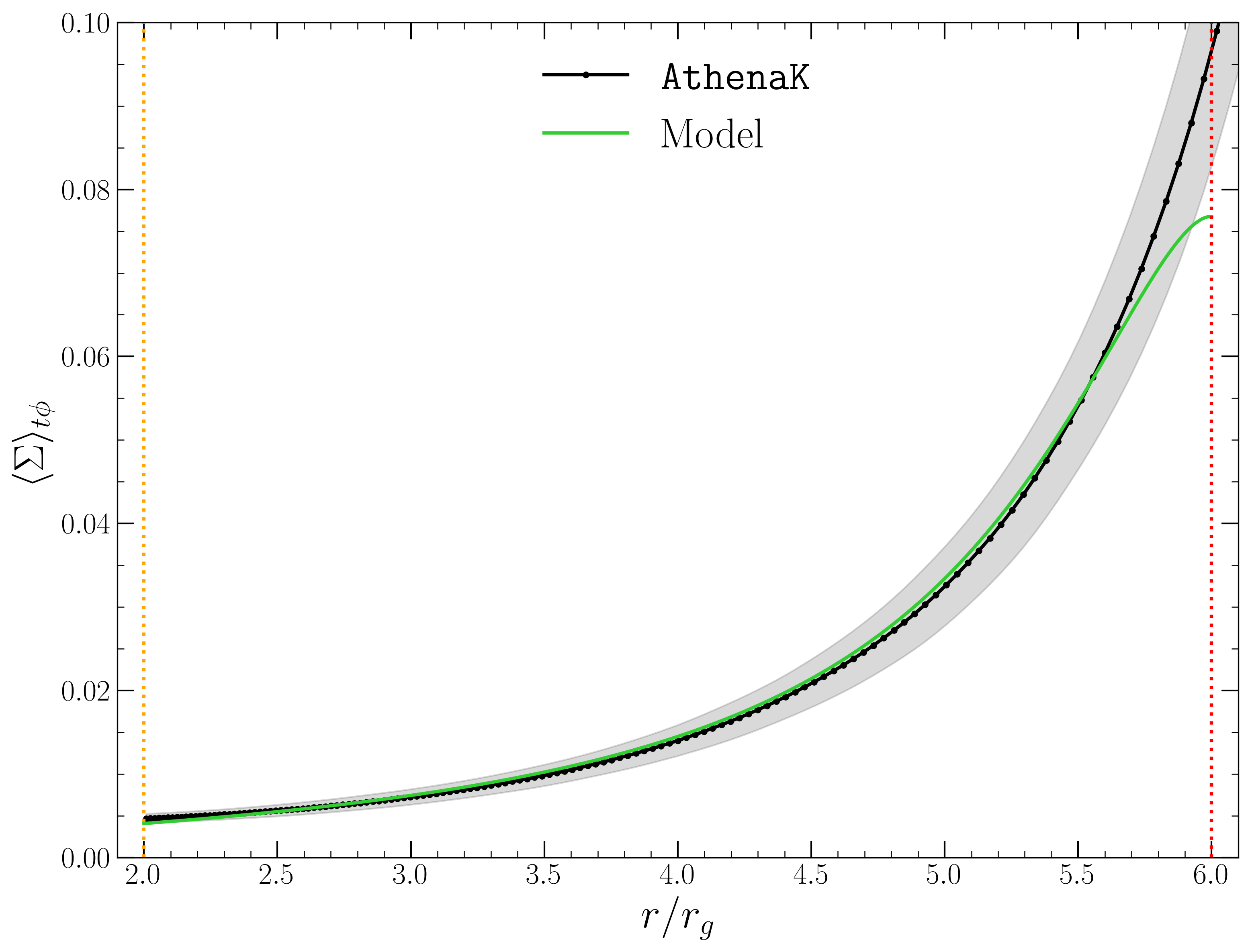}
    \includegraphics[width=0.49\linewidth]{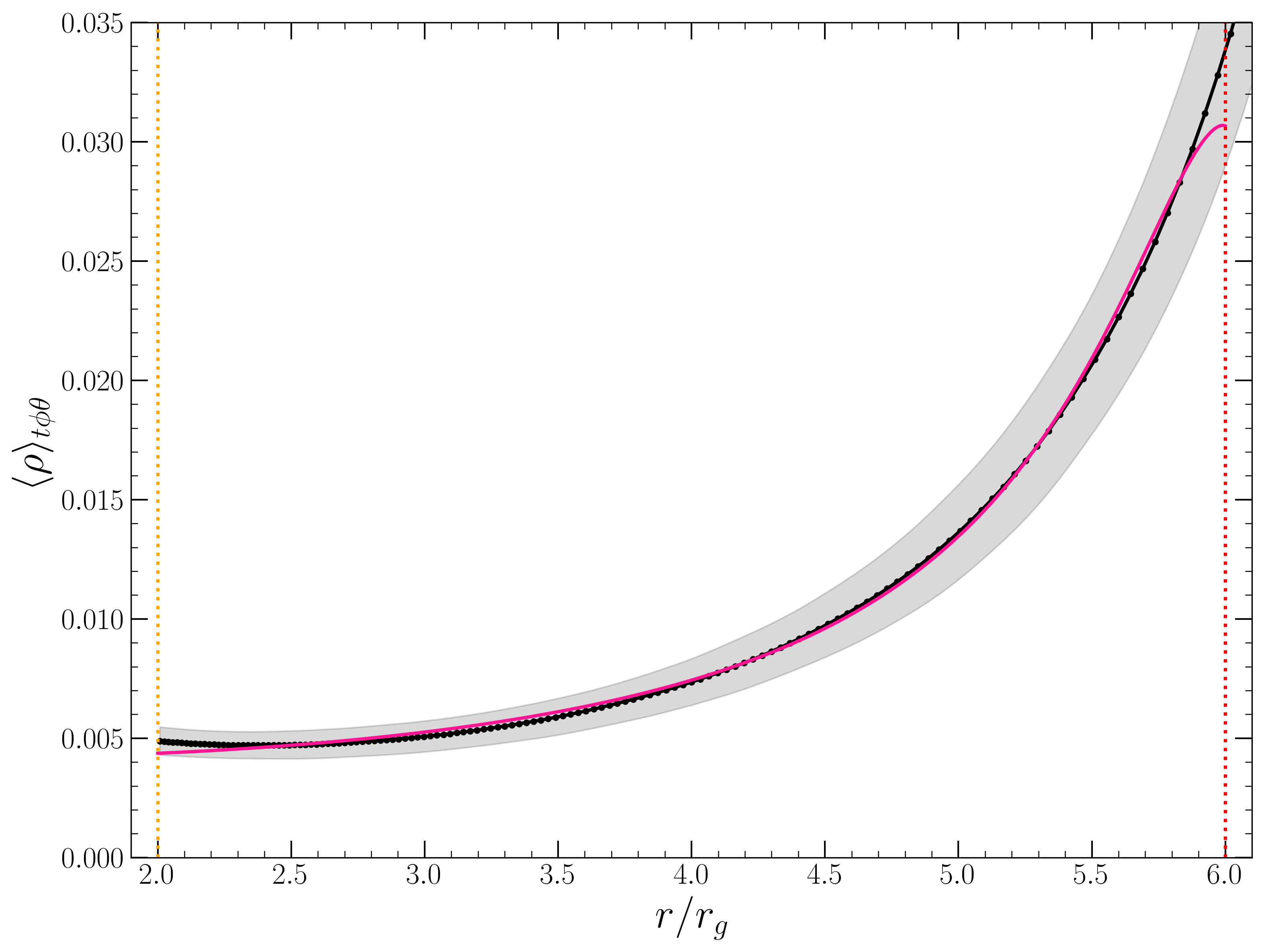}
    \includegraphics[width=0.49\linewidth]{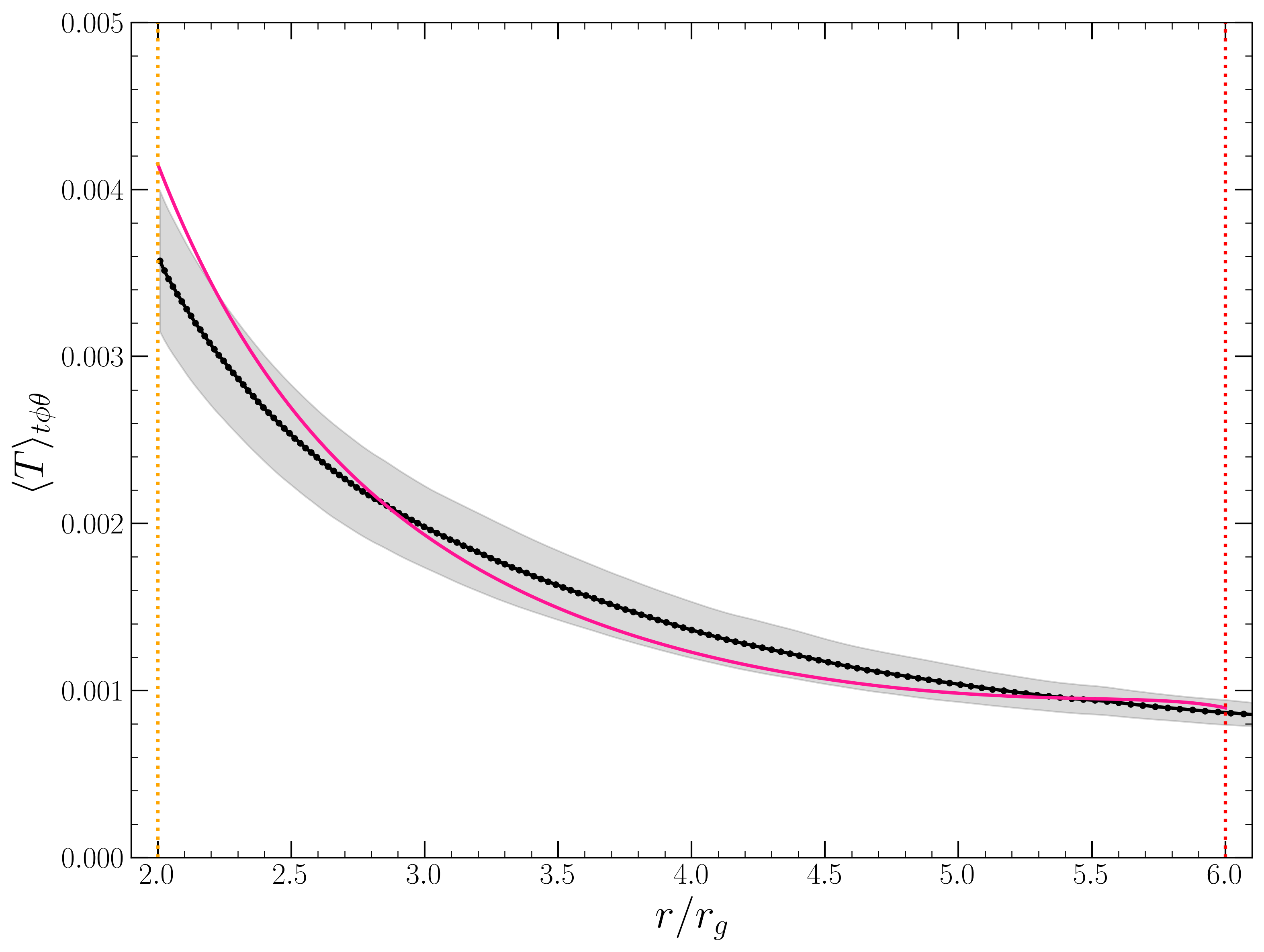}
    \includegraphics[width=0.49\linewidth]{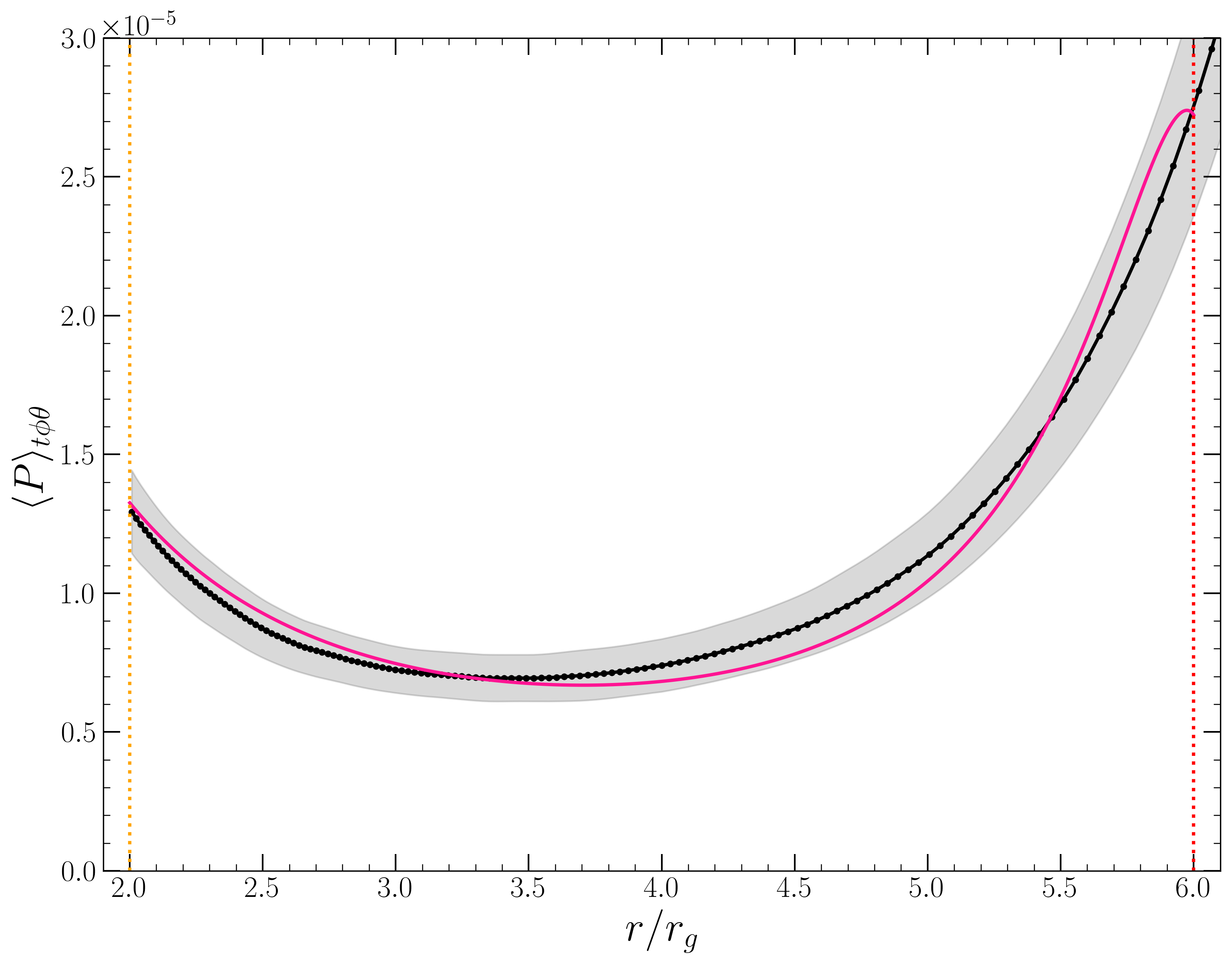}
    \caption{
    The plunging region radial velocity (top-left), specific entropy (top-right), surface density (mid-left), density (mid-right), temperature (bottom-left) and pressure (bottom-right) profiles for the lower resolution simulation, made in analogy with Figs.\,\ref{fig:urfit}\,,\,\ref{fig:EntropyandMagField}\,\&\,\ref{fig:thermofits}. Black dots show the temporally ($[12.5,17.5]\text{k}t_g$), azimuthally ($[0,2\pi]$) averaged simulated mid-plane quantities. The shaded regions represent $\pm1\sigma$ standard deviations. In the upper-right panel, the specific entropy represented by $K$, is modelled with a power law (pink). In the other panels, the pink lines show fitted MB23 models (Eqs.\,\ref{eq:densitymodel}, \ref{eq:pressuremodel} and \ref{eq:tempmodel}). For these, we assume that $K$ follows a power law and that the index $m$ is fixed to the value found from the $K$ profile. The surface density, density, pressure and temperature profiles are fitted separately so that they each have their own $\epsilon$ parameter (see Table\,\ref{tab:lowres_parameters}).
    }
    \label{fig:lowresfits}
\end{figure*}

Fig.\,\ref{fig:lowres_MdotandPhi} shows the evolution of the horizon mass accretion rate and the magnetic horizon flux as a function of time for the low resolution simulation. We observe that the magnetic horizon flux reaches a slightly lower level than in the high resolution case. Otherwise, the evolution is comparable. In Fig.\,\ref{fig:lowresfits}, we reproduce some of the profiles found in Figs.\,\ref{fig:urfit}\,,\,\ref{fig:EntropyandMagField}\,\&\,\ref{fig:thermofits} for the lower resolution simulation. The fitted parameters for these fits are found in Table\,\ref{tab:lowres_parameters}. Note that for these profiles, we time averaged between $12,500 \leq t/t_g \leq 17,500$. The profiles are very similar to those seen for the higher resolution simulation, in particular we still observe a sharp rise in the specific entropy, albeit with a slightly shallower power law index.
\begin{table}
\centering
\begin{tabular}{llll}
\hline
Refinement Level & $x,y \in $      & $z \in$ & Cell Spacing     \\ \hline
0     & $[-64,64]$ & $[-32,32]$ & $0.8 \, r_g/\text{cell}$ \\
1     & $[-64,64]$ & $[-16,16]$ &$0.4\,r_g/\text{cell}$ \\
2     & $[-32,32]$ & $[-8,8]$  & $0.2\,r_g/\text{cell}$ \\
3     & $[-16,16]$ & $[-4,4]$  &$0.1\,r_g/\text{cell}$ \\
4     & $[-8,8]$ & $[-2,2]$  &$0.05\,r_g/\text{cell}$ \\
\hline
\end{tabular}
\caption{
Static mesh refinement configuration for the lower resolution simulation.
}
\label{tab:lowres_refinement}
\end{table}

\begin{table}
\centering
\begin{tabular}{llll}
\hline
Model  & Quantity                                   & $\epsilon$ & $m$                  \\ \hline
Eq.\,\ref{eq:offsetgeo} & $\langle-U^r\rangle_{\rho, t \phi \theta}$ & $\it{0.043\pm0.004}$      & -                     \\
Power Law     & $\langle K \rangle_{t \phi \theta}$        & -          & $2.08$            \\
Eq.\,\ref{eq:surfdensitymodel}      & $\langle \Sigma \rangle_{t \phi}$          & $0.051$      & -              \\
Eq.\,\ref{eq:densitymodel}     & $\langle \rho \rangle_{t \phi \theta}$     & $0.031$      & $\ast$ 2.08          \\
Eq.\,\ref{eq:pressuremodel}     & $\langle P \rangle_{t \phi \theta}$        & $0.026$      & $\ast$ 2.08         \\
Eq.\,\ref{eq:tempmodel}    & $\langle T \rangle_{t \phi \theta}$        & $0.042$      & $\ast$ 2.08          \\\hline
\end{tabular}
\caption{
A summary of the same best-fit parameters as in Table\,\ref{tab:parameters}, for the lower resolution simulation.
}
\label{tab:lowres_parameters}
\end{table}


\bsp	
\label{lastpage}
\end{document}